\documentclass[review,12pt,sort&compress]{elsarticle}

\usepackage{epstopdf}                     

\usepackage{amssymb}
\usepackage{color}
\usepackage[left]{lineno}

\journal{Powder Technology}

\begin{document}

\begin{frontmatter}



\title{Plug regime in water fluidized beds in very narrow tubes \tnoteref{label_note_copyright} \tnoteref{label_note_doi}}

\tnotetext[label_note_copyright]{\copyright 2019. This manuscript version is made available under the CC-BY-NC-ND 4.0 license http://creativecommons.org/licenses/by-nc-nd/4.0/}


\tnotetext[label_note_doi]{Accepted Manuscript for Powder Technology,  v. 345, p. 234-246, 2019, DOI:10.1016/j.powtec.2019.01.009}


\author[label_david]{Fernando David C\'u\~nez Benalc\'azar}
\author[label_erick]{Erick de Moraes Franklin}

\address{School of Mechanical Engineering, University of Campinas - UNICAMP\\
Rua Mendeleyev, 200, Campinas - SP, CEP: 13083-860\\
Brazil}

\address[label_david]{e-mail: fernandodcb@fem.unicamp.br}
\address[label_erick]{phone: +55 19 35213375\\
e-mail: franklin@fem.unicamp.br, Corresponding Author}

\begin{abstract}
\begin{sloppypar}
The dynamics of granular plugs in water fluidized beds in narrow tubes is investigated here both experimentally and numerically. The fluidized beds were formed in a 25.4 mm-ID tube and consisted of alumina beads with 6 mm diameter and specific gravity of 3.69; therefore, the ratio between the tube and grain diameters was 4.23, which is considered a very narrow case. Three different beds were arranged, consisting of 250, 400 and 500 beads, and two water flow rates were imposed, corresponding, respectively, to superficial velocities of 0.137 and 0.164 m/s and to Reynolds numbers based on the tube diameter of 3481 and 4177. Under these conditions, it was possible to observe the formation of granular plugs that propagate with characteristic lengths and celerities. For the experiments, we filmed the fluidized bed with a high-speed camera, and automatically identified and tracked the plugs along images by using numerical scripts. For the numerical part, we performed simulations using a coupled CFD-DEM (computational fluid dynamics - discrete element method) code, together with numerical scripts to identify and track the granular plugs. We obtained, for the first time, the length scales and celerities of the granular plugs in very narrow tubes from experiments and numerical simulations, and the agreement between them is good.
\end{sloppypar}
\end{abstract}

\begin{keyword}
fluidized bed \sep water \sep narrow tubes \sep density waves \sep granular plugs
\\


\end{keyword}

\end{frontmatter}



\section{Introduction}
\label{intro}

A fluidized bed is a suspension of solid particles (grains) in a vertical tube submitted to an ascendant fluid flow. Given their high heat and mass transfers, fluidized beds are frequently employed in industrial processes such as, for example, combustion or gasification of coal and biomass, and drying, cooling and coating of solids. The dynamics of the fluidized grains is very rich, with different particle-fluid, particle-particle, and particle-walls interactions; therefore, the fluidized bed is susceptible to distinct patterns and flow regimes. Uniform fluidized beds are uncommon in industrial facilities, and instabilities usually appear. These instabilities give rise to transverse waves, bubbles, or long bubbles \cite{Nicolas, guazzelli2004fluidized}.

Over the past decades, many papers investigated the flow regimes in liquid fluidized beds \cite{Nicolas, Anderson2, Anderson, ElKaissy, Didwania, Zenit, Zenit2, Duru, Duru2, Aguilar, Ghatage}. Some of them were interested in fluidized beds in narrow ducts of rectangular cross section \cite{ElKaissy, Didwania, Duru}, where the thickness is between 10 to 100 grain diameters, and some few papers investigated liquid fluidized beds in narrow pipes \cite{Anderson, Zenit, Zenit2, Duru2, Aguilar, Ghatage}. Although there is not a consensus about it, we consider in this paper that for a narrow pipe the ratio between the tube and grain diameters is less than approximately 50. For liquid fluidized beds in narrow tubes, the dynamics and, consequently, the evolution of instabilities are different due to high confinement effects; therefore, this case is still to be understood.

Anderson and Jackson \cite{Anderson2} performed a stability analysis using a two-phase model. They showed that uniform fluidized beds are unstable and that density waves in the form of void regions may grow and propagate upwards in the bed. Because the growth of the instabilities is slower in the case of liquids, Anderson and Jackson \cite{Anderson} investigated experimentally the waves appearing in liquid fluidized beds in narrow tubes. They fluidized glass particles within 0.64 and 2.07 mm in tubes with diameters ranging from 12.7 to 38.1 mm. These sizes correspond, considering the tested conditions, to ratios between the tube and particle diameters from 10 to 29.5. The void waves measured by Anderson and Jackson \cite{Anderson} agreed well with the void disturbances predicted by stability analysis of \cite{Anderson2}.

El-Kaissy and Homsy \cite{ElKaissy} measured the initial waves that appear above minimum fluidization in liquid fluidized beds. The experiments were performed in a narrow duct 1.22 m-high of rectangular cross section (76.2 mm wide by 22.2 mm thick), and water was used to fluidize a granular bed consisting of beads with diameter ranging from 0.42 to 1.68 mm; therefore, the ratio of bed thickness to particle diameter was within 13 and 53. The appearance and propagation of waves were measured by flow visualization using light transmission. El-Kaissy and Homsy \cite{ElKaissy} found that two-dimensional transverse waves appear for fluid velocities just above the minimum fluidization velocity. The transverse waves are narrow regions of low particle fraction (void regions) that propagate upwards along the bed and persist up to some distance from the liquid distributor. For regions above this distance, coalescence and breakup occur and the void regions assume more complex shapes. The vertical region where the transverse waves exist decreases with the liquid velocity, and they disappear for fluid velocities larger than approximately twice the minimum fluidization velocity.

Didwania and Homsy \cite{Didwania} investigated the flow regimes and transitions in liquid fluidized beds. Their experiments were conducted in a 1.8 m-high column of rectangular cross section (300 mm wide by 31.5 mm thick) where glass beads with diameter within 0.25 and 1.19 mm and densities of 2990 and 3990 kg/m$^3$ were fluidized in water. The ratio of bed thickness to particle diameter was within 29 and 126, and the bed regimes were measured by using light transmission. Didwania and Homsy \cite{Didwania} identified four distinct regimes (in the order of increasing the fluid velocity): wavy flow, wavy flow with transverse structure, fine-scale turbulent flow, and bubbling regimes.

Duru et al. \cite{Duru2} measured the initial and saturated instabilities in solid-liquid fluidized beds in narrow tubes with ratio between the tube and grain diameters within 10 and 25. This range of diameters was chosen to achieve one-dimensional waves while avoiding arching effects. In order to measure the very first stages of the voidage instabilities, different forcing frequencies were imposed at the base of the bed by a piston mechanism. The authors found that the voidage disturbances reach a nonlinear saturated state after an initial growth. Duru et al. \cite{Duru2} present the first measurements of initial waves at their inception, and also the celerities and amplitudes of saturated waves.

The liquid-particle, particle-particle and particle-wall interactions are crucial to various phenomena in granular flows. Zenit et al. \cite{Zenit} measured the collisional particle pressure in liquid fluidized beds and gravity-driven flows. The authors used two different test sections, consisting of 50.8 and 106.1 mm-ID  tubes, where plastic, glass and steel beads of different diameters were submitted to different water flows. The ratio between the tube and particle diameters was between 8 and 52, and the particle Reynolds number based on the settling velocity was between 440 and 3665. A piezoelectric dynamic pressure transducer measured the granular pressure. In the specific case of liquid fluidized beds, Zenit et al. \cite{Zenit} quantified the granular pressure and found that it depends on the particle fraction in a non-monotonic way: it has a maximum value at particle fractions ranging from 30 to 40 $\%$. For lower particle fractions the collisions are less frequent, and for higher fractions the collisions have lower impulses due to reduced mobility. Comparisons with theoretical models showed that the best agreement is with the model proposed by Batchelor \cite{Batchelor}. The authors also found two contributions to the granular pressure: one from direct collisions of particles with the transducer and the other from bulk granular collisions transmitted through the liquid to the transducer.

For nearly the same cases presented in \cite{Zenit}, Zenit et al. \cite{Zenit2} measured the fluctuation component of grains in liquid fluidized beds and gravity-driven flows. The experimental setup was basically the same used in \cite{Zenit}, with the exception of the piezoelectric pressure transducer, which was not used. Instead, the authors used an impedance volume fraction meter, whose signal was sampled at 1 kHz. Zenit et al. \cite{Zenit2} found that the root mean square of the grains fluctuation varies non-monotonically with the particle fraction, reaching a maximum value for particle fractions between 30 and 45 $\%$, with a local maximum at 30 $\%$ and the absolute maximum at 45 $\%$. For a given solid fraction, the authors found that the fluctuations are higher for larger diameter particles. In addition, Zenit et al. \cite{Zenit2} found the coexistence of large-amplitude low-frequency and small-amplitude high-frequency fluctuations, the first being dominant for concentrations higher than 30 $\%$ and the latter for concentrations lower than 30 $\%$.

Aguilar-Corona et al. \cite{Aguilar} investigated the collision frequency and the coefficient of restitution of solid particles in a liquid fluidized bed. The authors used an 80 mm-ID tube filled with glass beads and a solution of water and Potassium Thiocyanate in order to match the optical index of the fluid with that of particles. The ratio between the tube and particle diameters was 13.33. The collision frequency and the particle-wall coefficient of restitution were measured based on images from both marked and unmarked particles. Within the ranges of Stokes and Reynolds numbers covered by their experimental setup, the authors found that the collision frequency is an increasing function of the concentration and that the normal restitution coefficient, computed based on the normal impact velocities, varies with the Stokes number. The value of the  restitution coefficient for liquid fluidized beds in narrow tubes obtained by \cite{Aguilar} is of great importance for numerical simulations.

To describe the dynamics of both liquid and particle phases, continuum and discrete models have been developed. The developments in the last decades increased the capabilities of CFD (computational fluid dynamics) simulations, which became popular for the study of two-phase flows. The detailed information about the local values of phase hold-ups, local flow velocities, and intermixing levels of the individual phases are difficult to obtain from experiments, but are readily obtained from simulations. Such information can be useful in understanding the transport phenomena in multiphase flows. DEM (discrete element method), which was first proposed by Cundall and Strack \cite{cundall1979discrete} using the soft-sphere collision model, has been used for a wide range of applications to determine the trajectories of particles by integrating the Newtonian equations of motion \cite{Kloss, Berger, Li}. The soft-sphere models allow for multiple particle overlap while the net contact force is obtained from the addition of all pair-wise interactions. In applications of gas-solid and liquid-solid flows, methods based on DEM coupled with CFD, known as CFD-DEM, are commonly used to predict the fluid flow and the behavior of solid particles \cite{Sun, Sun2, Liu, li2017modeling}. For these methods, the fluid is computed in an Eulerian grid while the solid particles are followed individually in a Lagrangian way.

Recently, Ghatage et al. \cite{Ghatage} investigated experimentally and numerically the transition from homogeneous to heterogeneous flow in liquid fluidized beds. In their experimental and numerical setups, the pipe diameter was 50 mm, the fluid was water and the particles were glass spheres with 5 mm diameter, so that the ratio between the tube and grain diameters was 10. For the experiments, they performed visualization of individual steel balls (intruders) and PIV (particle image velocimetry) employing optical index matching, with which they obtained the velocities of the liquid phase and the local void fraction. The authors did not measure directly the instabilities; instead, they found the critical condition for the transition from homogeneous to non-homogenous flow based on the settling velocity of the intruder. The numerical simulations were performed both with the two-fluid and CFD-DEM approaches using the commercial code Fluent. For the CFD-DEM, the authors had as limitation that the mesh size should be larger than the particle size, which made they employ a relatively coarse grid. The authors compared the experimental and numerical results with previous published stability models and the agreement was good. Ghatage et al. \cite{Ghatage} did not report explicitly the presence of granular plugs nor the corresponding lengths, celerities and frequencies.

In this paper we investigate one pattern appearing in beds fluidized by water in a very narrow tube. Confinement effects caused by the narrow tube combined with the viscous and virtual mass forces, which are significant under water, are the origin of alternating high- and low-compactness regions, known as plugs and bubbles, respectively. There have been very few studies about these patterns, and their dynamics is still to be understood. Few previous experimental studies were made to understand the problem, without exhausting it, and, to the authors' knowledge, no one investigated the very narrow case for which the ratio between the tube and grain diameters is between 2 and 5. In addition, no liquid-solid numerical simulation was conducted for this case. Successful numerical simulations of this problem would bring valuable data that are difficult to obtain experimentally. The objectives of this study are: (i) to understand the physics involved in the formation of granular plugs. This comprises the experimental measurements of wavelengths and celerities of granular plugs, and the use of CFD-DEM simulations, with all the pertinent mechanisms included in order to verify if the physics is well described by them. (ii) To validate a numerical methodology to simulate water fluidized beds that can be applied to different industrial scenarios. Both the experimental and numerical results are presented for the first time in this paper.

The fluidized beds were formed in a 25.4 mm-ID tube and consisted of alumina beads with 6 mm diameter and specific density of 3.69. The ratio between the tube and grain diameters was 4.23; therefore, we consider the tube as a very narrow one. For the experiments, we filmed the fluidized bed with a high-speed camera, and automatically identified and tracked the plugs along images by using numerical scripts. For the numerical part, we performed three-dimensional simulations using a coupled CFD-DEM code, together with numerical scripts to identify and track the granular plugs. The fluid flow is computed with the open source code OpenFOAM, based on FVM (finite volume method), while the granular dynamics is computed with the open source code LIGGGHTS \cite{Kloss, Berger}, which is based on DEM, and both are linked via the open source code CFDEM \cite{Goniva}. Our numerical results using CFD-DEM show the formation of plugs under water. We obtained, for the first time, the length scales and celerities of the granular plugs in very narrow tubes from both the experiments and the numerical simulations, and the agreement between them is good. These results are encouraging, since the numerical methodology used can be applied to more complex industrial problems.

The next sections present the experimental setup, model description, numerical setup, and experimental and numerical results. The following section presents the conclusions.

\section{Experimental setup}

The experimental setup consisted of a water reservoir, a heat exchanger, a centrifugal pump, a flow meter, a flow homogenizer, a 25.4 mm-ID tube with vertical and horizontal sections, and a return line. The centrifugal pump was able to impose water flow rates from 0 to 4100 l/h, and water flowed in a closed loop in the order just described. The flow rates were adjusted by controlling the pump rotation together with a set of valves.

\begin{figure}[ht]
	\centering
	\includegraphics[width=0.8\columnwidth]{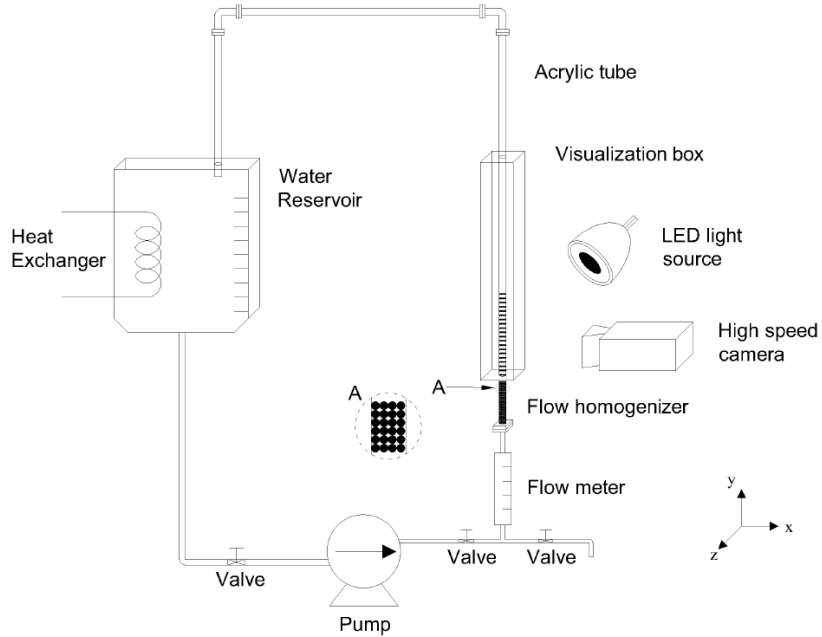}
	\caption{Layout of the experimental setup.}
	\label{fig:1}
\end{figure}

\begin{figure}[ht]
	\centering
	\includegraphics[width=0.3\columnwidth]{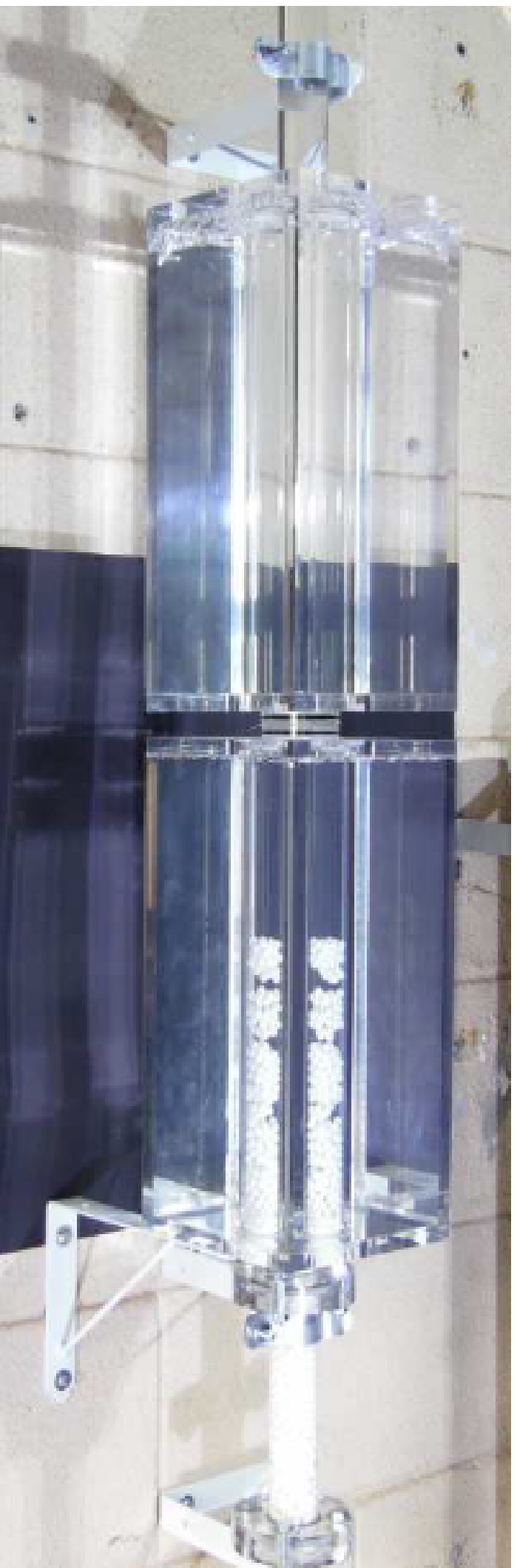}
	\caption{Test section.}
	\label{fig_test_section}
\end{figure}

The vertical section was a 1.2 m-long, 25.4 mm-ID transparent PMMA (Polymethyl methacrylate) tube, of which 0.65 m was the test section. The vertical tube was vertically aligned within $\pm 3^{\circ}$ and, in order to avoid parallax distortions, a visual box filled with water was placed around the test section. The flow homogenizer consisted of a 150 mm-long tube containing packed alumina beads with $d$ = 6 mm placed between fine wire screens, where $d$ is the bead diameter, and it was used to assure uniform water flows at the inlet of the test section. The function of the heat exchanger was to assure water temperatures within 25$^{\circ}$C $\pm$ 3$^{\circ}$C. Fig.\ref{fig:1} shows the layout of the experimental apparatus and Fig. \ref{fig_test_section} shows a photograph of the test section.

Three different beds were arranged, consisting of 250, 400 and 500 alumina beads with $d$ = 6 mm $\pm$ 0.03 mm and $S$ = $\rho_p / \rho_f$ = 3.69, where $\rho_p$ is the density of the bead material and $\rho_f$ the density of the fluid. The ratio between the tube, $D$, and the mean grain diameter was $D/d$ = 4.23, and the Stokes and Reynolds numbers based on terminal velocities were $St_t \,=\, v_t d \rho_p / (9\mu_f)$ = 1650 and $Re_t \,=\, \rho_f  v_t d / \mu_f$ = 4026, respectively, where $v_t$ is the terminal velocity of one single particle and $\mu_f$ is the dynamic viscosity of the fluid. The bed heights at the inception of fluidization $h_{mf}$ were in average 117, 181 and 225 mm for the beds consisting of 250, 400 and 500 particles, respectively, from which the liquid volume fraction at the inception of fluidization $\varepsilon_{mf}$ was computed. The settling velocity at the inception of fluidization, computed based on the Richardson--Zaki correlation, was $v_s = v_t\varepsilon_{mf}^{2.4}$ = 0.13 m/s. Two water flow rates were imposed, $Q$ = 250 l/h and $Q$ = 300 l/h, for which the corresponding superficial velocities $\overline{U} = 4Q/\pi D^2$, fluid velocities through the packed bed $U_f=\overline{U} / \varepsilon_{mf}$, Reynolds numbers based on the tube diameter $Re_D \,=\, \rho_f  \overline{U} D / \mu_f$, and Reynolds numbers based on the grain diameter $Re_d \,=\, \rho_f  \overline{U} d / \mu_f$ are summarized in Tab. \ref{table:table1}.

\begin{table}[ht]
\caption{Grain diameter $d$, terminal Reynolds number $Re_t$, terminal Stokes number $St_t$, water flow rate $Q$, superficial velocity $\overline{U}$, Reynolds number based on the tube diameter $Re_D$, Reynolds number based on the grain diameter $Re_d \,=\, \rho_f  \overline{U} d / \mu_f$, settling velocity $v_s$, and fluid velocities trough the packed bed $U_f$.}
\label{table:table1}
\centering
\begin{tabular}{c c c c c c c c c}  
\hline\hline
$d$ & $Re_t$  & $St_t$ & $Q$ & $\overline{U}$ & $Re_D$ & $Re_d$ & $v_s$ & $U_f$\\
mm & $\cdots$ & $\cdots$ & $l/h$ & $m/s$ & $\cdots$ & $\cdots$ & $m/s$ & $m/s$\\ [0.5ex] 
\hline 
6 & 4026 & 1650 & 250 & 0.137 & 3481 & 822 & 0.13 & 0.27\\
6 & 4026 & 1650 & 300 & 0.164 & 4177 & 987 & 0.13 & 0.32\\
\hline
\hline 
\end{tabular}
\end{table}

The fluidized bed was filmed with a high-speed camera of CMOS (Complementary Metal Oxide Semiconductor) type having a resolution of 1600 $px$ $\times$ 2560 $px$ at frequencies up to 1400 Hz at full resolution. In order to provide the necessary light for low exposure times while avoiding beating between the light source and the camera frequency, LED (low emission diode) lamps were branched to a continuous current source. Given the tubular shape of the test section, the ROI (region of interest) was fixed to 248 $px$ $\times$ 2400 $px$, and the total fields used measured 42 mm $\times$ 406 mm, corresponding to areas of 35 px $\times$ 35 px for each grain and around 140 px $\times$ 350 px for each plug. In all the experiments, the camera frequency was set to between 100 Hz and 200 Hz. The number of acquired images for each test was 5000 and the total number of tests was 6, giving a total of 30000 images to be analyzed. Numerical codes were written in order to process the obtained images, their description can be found in \cite{Alvarez}.

Prior to each test, a water flow corresponding to $\overline{U}$ = 0.55 m/s was imposed during 1 to 5 s in order to suspend all the grains, which settled afterwards and formed a static bed. When all the particles were settled, the desired flow rate was fixed and the test run began.

\section{Model description}
\label{section_model}

The CFD-DEM approach is relatively well documented in the literature; therefore, we delineate in this paper the main features of the model equations that are fundamental to the present problem.

\subsection{Liquid phase}
The fluid phase is described by the averaged incompressible Navier-Stokes equations for two-phase flows. Assuming constant fluid density, the mass and momentum equations for the fluid are, respectively:

\begin{equation}
{\frac{\partial{\rho_{f}\varepsilon_{f}}}{\partial{t}}+\nabla\cdot(\rho_{f}\varepsilon_{f}\vec{u}_{f})=0}
\label{mass}
\end{equation}

\begin{equation}
\frac{\partial{\rho_{f}\varepsilon_{f}\vec{u}_{f}}}{\partial{t}} + \nabla \cdot (\rho_{f}\varepsilon_{f}\vec{u}_{f}\vec{u}_{f}) = -\varepsilon_{f}\nabla P + \varepsilon_{f}\nabla\cdot \vec{\vec{\tau}}_{f} + \varepsilon_{f}\rho_{f}\vec{g}+\vec{F}_{pf}
\label{qdm}
\end{equation}

\noindent where $P$ is the pressure, $\vec{\vec{\tau}}_{f}$ is the stress tensor, $\vec{g}$ is the acceleration of gravity, and $\vec{u}_{f}$ and $\varepsilon_{f}$ represent the mean velocity and volume fraction of the fluid phase, respectively. The term $\vec{F_{pf}}$ corresponds to the momentum transfer between the fluid and the solid particles, and it can be computed by Eq. \ref{Fpf} \cite{li2017modeling},

\begin{equation}
{\vec{F}_{pf}=\frac{1}{V_{cell}}\sum_{\forall p \in cell}\frac{V_{p}\beta}{1-\varepsilon_{f}}\left(\vec{u}_{p}-\vec{u}_{fp}\right)}
\label{Fpf}
\end{equation}

\noindent where $V_{cell}$ and $V_{p}$ are the volumes of the considered cell and particle, respectively, $\beta$ is the coefficient of momentum transfer between phases (due to the drag force), $\vec{u}_{p}$ is the particle velocity, and $\vec{u}_{fp}$ is the liquid velocity at the particle position. The latter is usually obtained by interpolation.

\subsection{Particle motion}

The translational and rotational motion of each particle is calculated based on Newton's second law. The linear and angular momentum equations are given, respectively, by Eqs. \ref{Fp} and \ref{Tp},

\begin{equation}
m_{p}\frac{d\vec{u}_{p}}{dt}=-V_{p}\nabla P+\vec{F}_{fp} + \vec{F}_{visc} + \vec{F}_{vm}+m_{p}\vec{g}+\sum_{i\neq j}^{N_c} \left(\vec{F}_{c,ij} \right) + \sum_{i}^{N_w} \left( \vec{F}_{c,iw} \right)
\label{Fp}
\end{equation}

\begin{equation}
I_{p}\frac{d\vec{\omega}_{p}}{dt}=\sum_{i\neq j}^{N_c} \vec{T}_{p,ij} + \sum_{i}^{N_w} \vec{T}_{p,iw}
\label{Tp}
\end{equation}

\noindent where $m_{p}$ is the mass of a particle, $\vec{T}_{p}$ is the torque arising from the tangential component of the contact force, and $I_{p}$ and $\vec{\omega}_{p}$ are the moment of inertia and angular velocity of a particle, respectively. The terms on the right-hand side of Eq. \ref{Fp} represent the forces generated by the liquid pressure gradients, drag exerted by the fluid (classical drag, described below), viscous flow in the vicinity of particles, virtual mass, gravity, and solid contact between particles and between particles and the tube wall, respectively. The term $\vec{T}_{p,ij}$ represents the torque generated by the tangential component of the contact force between particles $i$ and $j$, and $\vec{T}_{p,iw}$ the torque generated by the tangential component of the contact force between particle $i$ and the wall. The inter-particle forces and torques are summed over the $N_c$ - 1 particles in contact with particle $i$, where $N_c$ is the total number of particles in contact. The particle-wall forces and torques are summed over the $N_w$ particles in contact with the wall. The contact forces between particles and between particles and the wall are calculated based on the soft-particle method \cite{cundall1979discrete}.

The drag force $\vec{F}_{fp}$, given by Eq. \ref{Fpf}, is determined for each particle. The drag force depends not only on the relative velocity between the solid particle and fluid, but also on the presence of neighboring particles, i.e., the local volume fraction of the solid phase. This is taken into account by the drag correlation coefficient $\beta$, which considers the presence of other particles. In this study, we consider that $\beta$ is given by the Gidaspow model \cite{gidaspow1994multiphase},

\begin{equation}
\beta=\left\{
\begin{array}{cc}
\frac{3}{4}C_{d}\frac{\rho_{f}\varepsilon_{f}\left(1-\varepsilon_{f}\right)\left|\vec{u}_{f}-\vec{u}_{p}\right|}{d}\varepsilon_{f}^{-2.65} & \left(1-\varepsilon{f}\right) \leq 0.2\\
\\
150\frac{\left(\varepsilon_{f}\right)^{2}\vec{u}_{f}}{\varepsilon_{f}d^{2}}+1.75\frac{\rho_{f}\left(1-\varepsilon_{f}\right)\left|\vec{u}_{f}-\vec{u}_{p}\right|}{d} & \left(1-\varepsilon{f}\right) > 0.2\\
\end{array}\right.
\label{beta}
\end{equation}

\noindent with,

\begin{equation}
C_{d}=\left(0.63+0.48\sqrt{V_{r}/Re}\right)
\label{cd}
\end{equation}

\noindent where the particle Reynolds number $Re$ is defined as

\begin{equation}
Re=\frac{\rho_{f}d\left|\vec{u}_{f}-\vec{u}_{p}\right|}{\mu_{f}}
\label{Re}
\end{equation}

\noindent and $V_{r}$ is given by Eq. \ref{Vr}

\begin{equation}
V_{r}=0.5\left(A_{1}-0.06Re+\sqrt{\left(0.06\right)^{2}+0.12Re\left(2A_{2}-A_{1}+A_{1}^{2}\right)}\right)
\label{Vr}
\end{equation}

\noindent with,

\begin{equation}
\begin{array}{cc}
A_{1}=\varepsilon_{f}^{4.14} \\
\\
A_{2}=\left\{
\begin{array}{cc}
0.8\varepsilon_{f}^{1.28} & \varepsilon_{f} \leq 0.85\\
\\
\varepsilon_{f}^{2.65} & \varepsilon_{f} > 0.85
\end{array}\right.
\end{array}
\label{A}
\end{equation}

In the case of liquids, in addition to drag correlations used in the present work, viscous forces must be accounted for. In our simulations, the viscous forces are computed from velocity gradients around the solid particles. Finally, when a solid is accelerated through a fluid, part of the fluid is accelerated at the expense of work done by the solid. This additional work is associated to the virtual mass force, $\vec{F}_{vm}$, given by Eq. \ref{Fvm}. The virtual mass force acting on a particle, usually neglected in the case of gases, must be considered in the case of liquids.

\begin{equation}
\vec{F}_{vm}=-0.5\left(1-\varepsilon_{f}\right)\rho_{f}\left(\frac{d\vec{u}_{f}}{dt}-\frac{d\vec{u}_{p}}{dt}\right)
\label{Fvm}
\end{equation}

\subsubsection{Particle-particle and particle-wall interactions}

Usually, the contact force is divided into the normal, $\vec{F}_{cn}$, and tangential, $\vec{F}_{ct}$, components, as shown in Eq.\ref{Fc}. We use here the Cundall and Strack’s model \cite{cundall1979discrete}, consisting of a spring, a dash-pot and a slider to model the contact forces. The model then makes use of stiffness $k$, damping $\eta$, and friction $\mu_{fr}$ coefficients, which are related to deformation distances in normal and tangential directions, $\delta_n$ and $\delta_t$, respectively. For a HSD (Hertzian spring-dashpot) model, the normal and tangential components are given, respectively, by Eqs. \ref{Fcn} and \ref{Fct}.

\begin{equation}
\vec{F}_{c,ij}=\vec{F}_{cn,ij}+\vec{F}_{ct,ij}
\label{Fc}
\end{equation}

\begin{equation}
\vec{F}_{cn,ij}=\left(-k_{n}\delta_{nij}^{3/2}-\eta_{n}\vec{u}_{ij}\cdot\vec{n}_{ij}\right)\vec{n}_{ij}
\label{Fcn}
\end{equation}

\begin{equation}
\vec{F}_{ct,ij}= \left( -k_{t}\delta_{tij} - \eta_{t}\vec{u}_{sij}\cdot\vec{t}_{ij} \right) \vec{t}_{ij}
\label{Fct}
\end{equation}

The friction coefficient is usually obtained empirically. We obtained the values of the stiffness $k$, damping $\eta$ and friction $\mu_{fr}$ coefficients from Tsuji et al. \cite{tsuji1992lagrangian}. In Eqs. \ref{Fcn} and \ref{Fct}, $\vec{n}_{ij}$ is the unit vector connecting the centers of particles with direction from $i$ to $j$, $\vec{t}_{ij} = \vec{u}_{sij}/ \left| \vec{u}_{sij} \right|$, $k_{n}$ and $k_{t}$ are the normal and tangential stiffness coefficients, respectively, $\eta_{n}$ and $\eta_{t}$ are the normal and tangential damping coefficients, respectively, $\vec{u}_{ij}$ is the relative velocity between particle $i$ and particle $j$, and $\vec{u}_{sij}$ is the slip velocity at the contact point. The latter is computed by Eq. \ref{usij}.

\begin{equation}
\vec{u}_{sij}=\vec{u}_{ij}-\left(\vec{u}_{ij}\cdot\vec{n}_{ij}\right)\vec{n}_{ij}+\left( |\vec{r}_{i}| \vec{\omega}_{i} + |\vec{r}_{j}| \vec{\omega}_{j}\right)\times \vec{n}_{ij}
\label{usij}
\end{equation}

For the Hertzian model, $k_{n}$ and $k_{t}$ are given by Eqs.\ref{kn} and \ref{kt} \cite{tsuji1992lagrangian},

\begin{equation}
k_{n}=\frac{4}{3}\sqrt{r_{eff}}\frac{E}{2\left(1-\sigma_{j}^{2}\right)}
\label{kn}
\end{equation}

\begin{equation}
k_{t}=8\sqrt{r_{eff}\delta_{nij}}\frac{G}{2\left(2-\sigma\right)}
\label{kt}
\end{equation}

\noindent where $E$ is the Young's modulus, $\sigma$ is the Poisson ratio, $G=E/(2(1+\sigma))$, and $r_{eff}=r_{i}r_{j}/(r_{i}+r_{j})$ is the reduced radius. The damping coefficient in the normal direction with the prefactor proposed by \cite{li2017modeling} is given by Eq.\ref{nn},

\begin{equation}
\eta_{n}= \left[ -0.719\ln\left(\left(e+0.08\right)/1.078\right) \right] \left(m_{eff}k_{n}\right)^{1/2}\delta_{n}^{1/4}
\label{nn}
\end{equation}

\noindent where $m_{eff}=m_{i}m_{j}/(m_{i}+m_{j})$ is the reduced mass. In the present study, we consider the damping coefficient in the tangential direction $\eta_{t}$ equal to $\eta_{n}$ \cite{tsuji1992lagrangian}.

Finally, if Eq. \ref{relfc} is valid, then sliding occurs and the tangential force is calculated by Eq.\ref{fct}

\begin{equation}
\left|\vec{F}_{ct,ij}\right|>\mu_{fr}\left|\vec{F}_{cn,ij}\right|
\label{relfc}
\end{equation}

\begin{equation}
\vec{F}_{ct,ij}=-\mu_{fr}\left|\vec{F}_{cn,ij}\right|\vec{t}_{ij}
\label{fct}
\end{equation}

The same approach is used for particle-wall contacts by replacing the index $j$ by $w$.

\section{Numerical setup}
\label{section_numerical}

The parameters used in the numerical simulations are the same as used in the experiments. We considered a 0.45 m-long and 25.4 mm-ID vertical tube, filled with alumina spheres with $d$ = 6 mm and water. Three different beds were arranged, consisting of 250, 400 and 500 alumina beads, and water flows corresponding to cross-sectional mean velocities of $\overline{U}$ = 0.137 and 0.164 m/s were imposed at the tube inlet.

The main issue to implement this problem in standard CFD-DEM codes is the limitation concerning the mesh size, which has to be larger than the size of particles. In the present case, the ratio between the tube and grain diameters is approximately 4, so that the mesh size limitation is not acceptable. In order to prevent it, we employed a big particle void fraction model (www.cfdem.com), which is used when the particle is larger than the CFD cells. A given value of void fraction is associated to the cells whose centers are inside the particle: equal to 1 for the fluid and 0 for the solid. To use the so determined void fractions in Eqs. \ref{mass} and \ref{qdm}, CFDEM artificially treats the grains as porous particles by increasing their volume while maintaining the original volume of the solid phase \citep{Kloss2, Mondal}. This artificial method allows the CFD-DEM code to use Eqs. \ref{mass} and \ref{qdm} with cells smaller than the particle size.

\subsection{Computational geometry}

\begin{figure}[ht]
	\centering
	\includegraphics[width=7cm,clip]{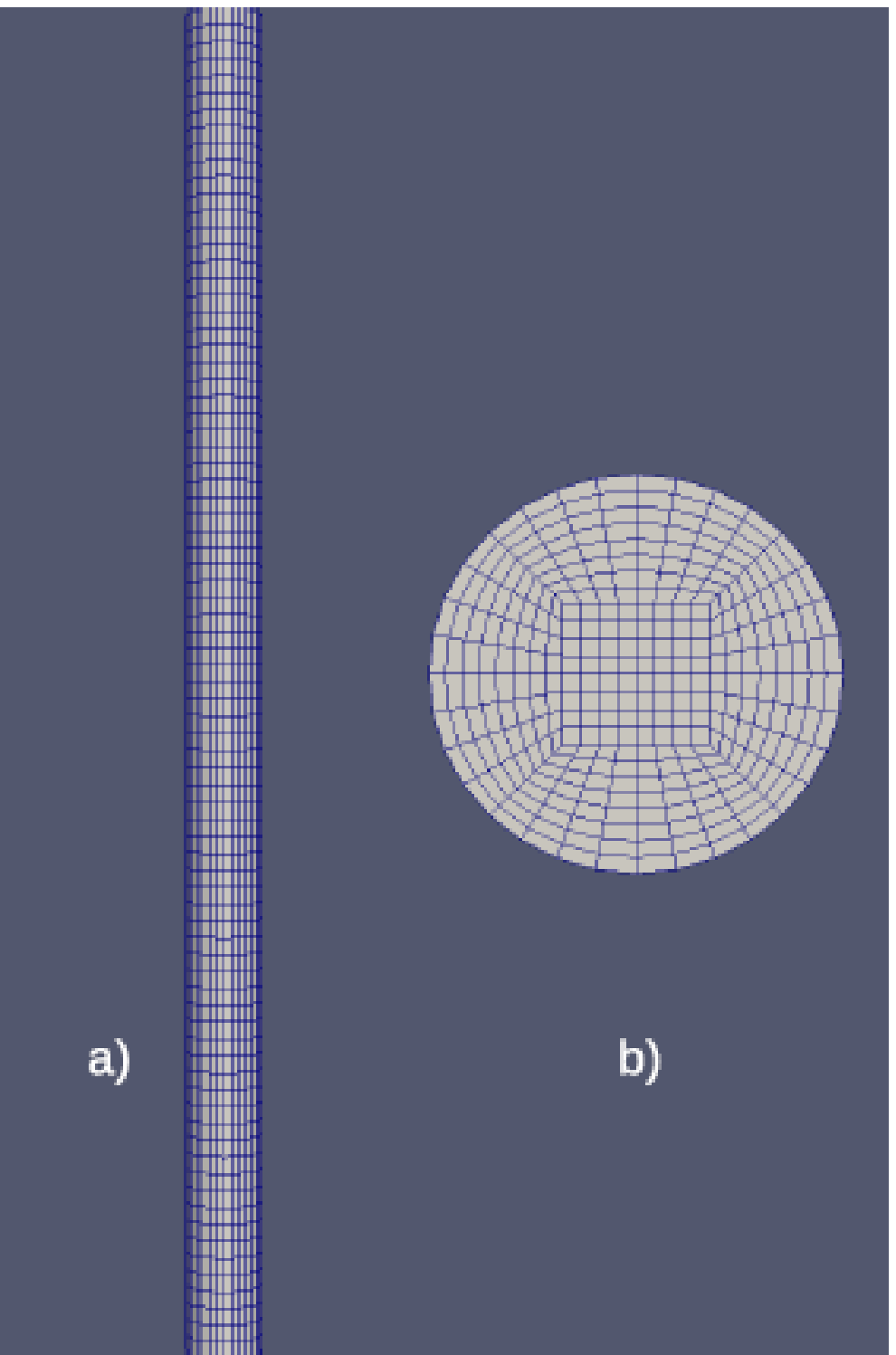}
	\caption{Computational geometry: (a) Side view; (b) Bottom view.}
	\label{mesh}  
\end{figure}

A 3D geometry was created using the mesh generation utility BlockMesh, which is included in OpenFOAM, to simulate the 3D water fluidized bed in an Eulerian-Lagrangian approach. We generated a hexahedral mesh with a total number of 25600 cells. The computational geometry is showed in Fig.\ref{mesh}.

\subsection{Boundary conditions and numerical methods}

\begin{figure}[ht]
	\centering
	\includegraphics[width=7cm,clip]{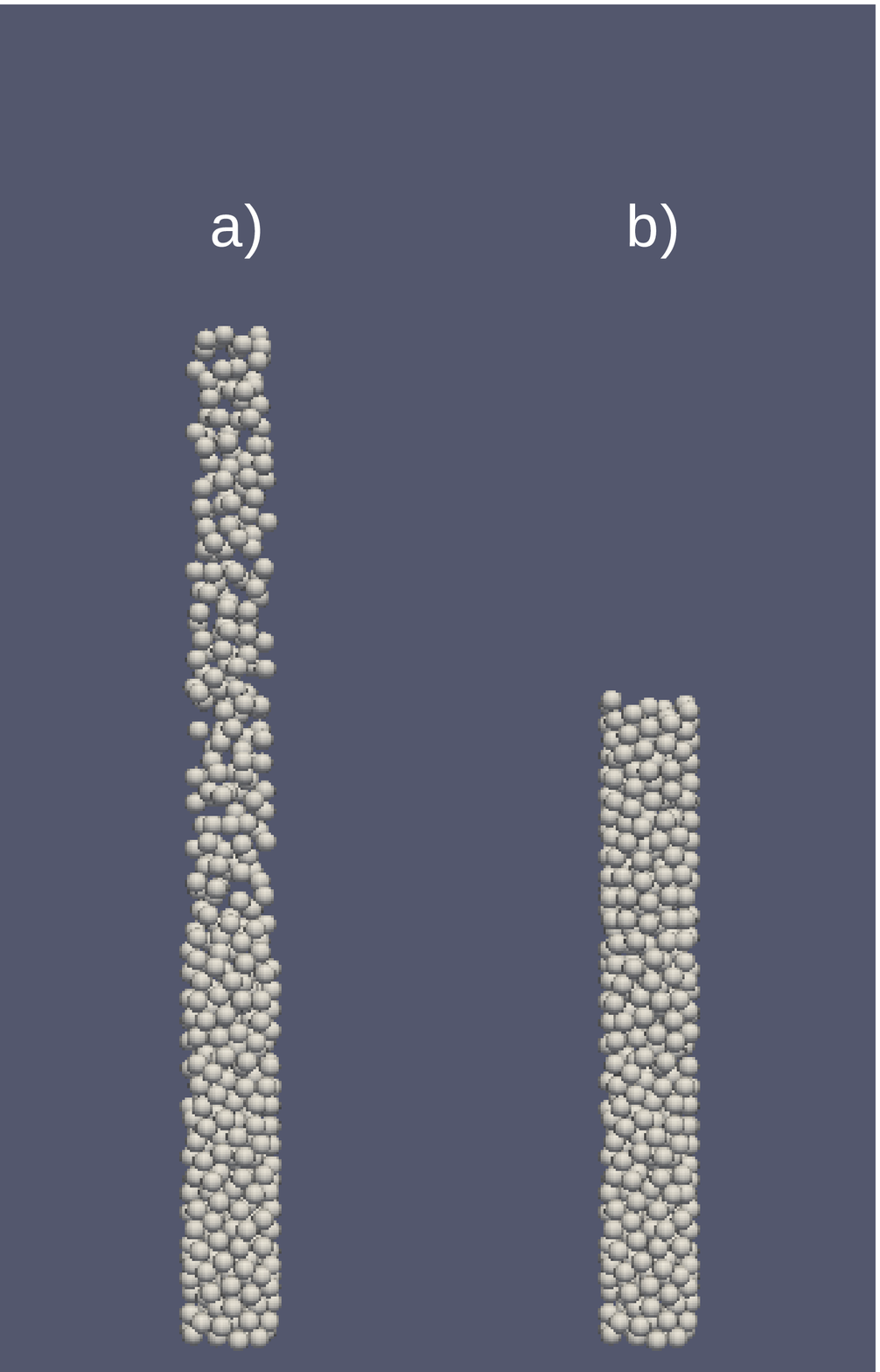}
	\caption{Particles positions: (a) falling particles that were randomly distributed at the initial condition; (b) relaxed state, at the end of initialization.}
	\label{part}  
\end{figure}

The calculations of liquid flow and particle motion are three-dimensional. Initially, the particles are allowed to fall freely under water in the absence of a water flow. After some time for energy dissipation through inelastic collisions, the bed reaches an almost stagnant state, as shown in Fig.\ref{part}. The water flow is not turned on until the initialization is completed.

The lateral walls are impenetrable, with no slip condition for the fluid, so that the tangential and normal velocities of the liquid phase are set to zero. At the bottom boundary, the liquid velocity in the vertical direction is set equal to the superficial velocity $\overline{U}$, and the velocity of particles is zero. At the top of the bed, the liquid pressure is specified, and the velocity gradient of the liquid phase is assumed to be equal to zero. 

Numerical simulations were performed by means of CFDEM \cite{Goniva}, an open source CFD-DEM code that couples OpenFOAM to LIGGGHTS. The equations of fluid motion were solved by the finite volume method with the use of the numerical method PIMPLE (merged PISO and SIMPLE algorithms, www.openfoam.com). The differential equations of fluid motion were considered in an Eulerian framework in which the fluid cells are fixed to a reference frame, whereas the equations for the solid particles were considered in a Lagrangian framework. The motion equations of solid particles, with the mutual interaction between fluid and particles taken into account, were simultaneously solved with the motion equations of the fluid to provide particle positions, particle velocities, and fluid velocities. The particle fraction in the cell was obtained from the particle positions.

In the present simulations, a constant time step of $5.0\times 10^{-4}s$ was used for liquid phase and a time step of $1.0\times 10^{-5}s$ was used for the solid particles. The computational parameters are listed in Table \ref{tabsim}, where $N$ is the total number of solid particles in the bed.

\begin{table}[ht]
	\begin{center}
	\caption{Parameters used in simulations}
	\begin{tabular}{c c}
		\hline\hline
		Particle diameter $d$ (mm)  & 6  \\
		Particle density $\rho_p$ (kg/m$^3$) & 3690 \\
		Number of particles $N$ & 250, 400 and 500 \\
		Young's Modulus $E$ (GPa) & 300   \\
		Poisson ratio $\sigma$ & 0.21 \\
		Restitution coefficient $e$ & 0.5\\
		Friction coefficient $\mu_{fr}$ & 0.5\\
		Liquid density $\rho_f$ (kg/m$^3$) & 1000  \\
		Liquid viscosity $\mu_f$ (Pa.s) & 1$\times 10^{-3}$\\
		\hline
	\end{tabular}
		\label{tabsim}
    \end{center}
\end{table}

The coefficient of restitution was considered as approximately 0.5 based on the experimental results of Aguilar-Corona et al. \cite{Aguilar}, and the Young's modulus, Poisson ratio and friction coefficient were obtained from \cite{tsuji1992lagrangian}.

\section{Results}

\subsection{Experiments}

Under the tested conditions, granular plugs and liquid bubbles (void regions) occupying the entire tube cross section were observed in the fluidized bed. Those forms, that were nearly one dimensional, propagated upward with characteristic lengths and celerities. They result from a competition between the mechanisms that tend to displace the particles along the vertical direction, and those that tend to block them over the pipe cross section. The former are associated with the fluid drag and viscous forces, which entrain the grains upward, and gravity, which  entrains the grains downward. As for any fluidized bed, their action maintains the grains suspended, which can oscillate and move over distances that depend on the bed conditions. The latter mechanisms are related to the formation of arches. Because this is a very narrow case, the probability for the formation of arches is relatively high; therefore, considering the friction between grains and between grains and the walls, a part of vertical forces is redirected toward the horizontal direction. The redirection of forces limits the action of entraining forces, creating blockage regions. In addition, for ensembles of grains longer than one tube diameter, Janssen effect is expected.

\begin{figure}[ht]
 	\begin{minipage}{0.5\linewidth}
 		\begin{tabular}{c}
 			\includegraphics[width=0.90\linewidth]{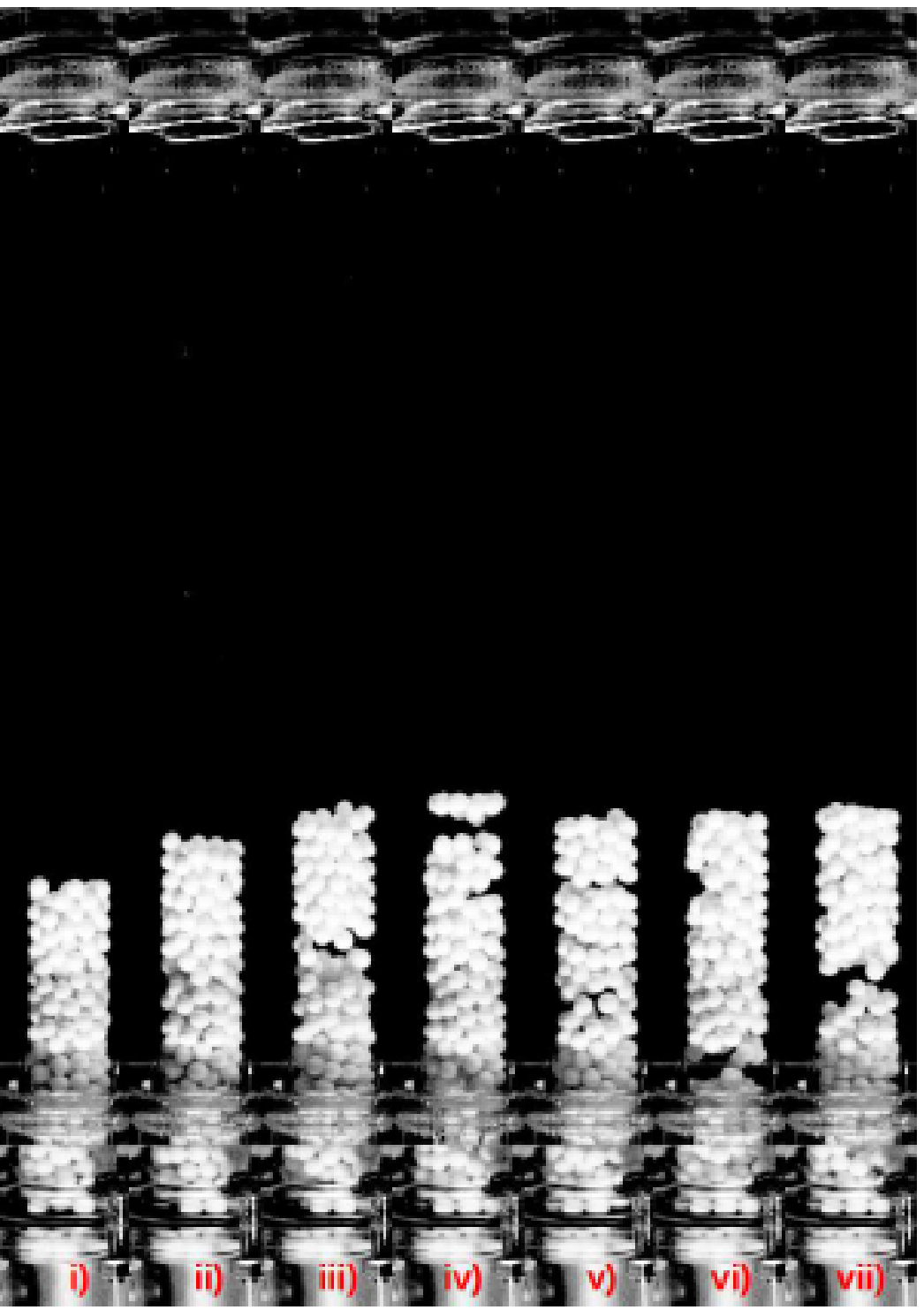}\\
 			(a)
 		\end{tabular}
 	\end{minipage}
 	\hfill
 	\begin{minipage}{0.5\linewidth}
 		\begin{tabular}{c}
 			\includegraphics[width=0.90\linewidth]{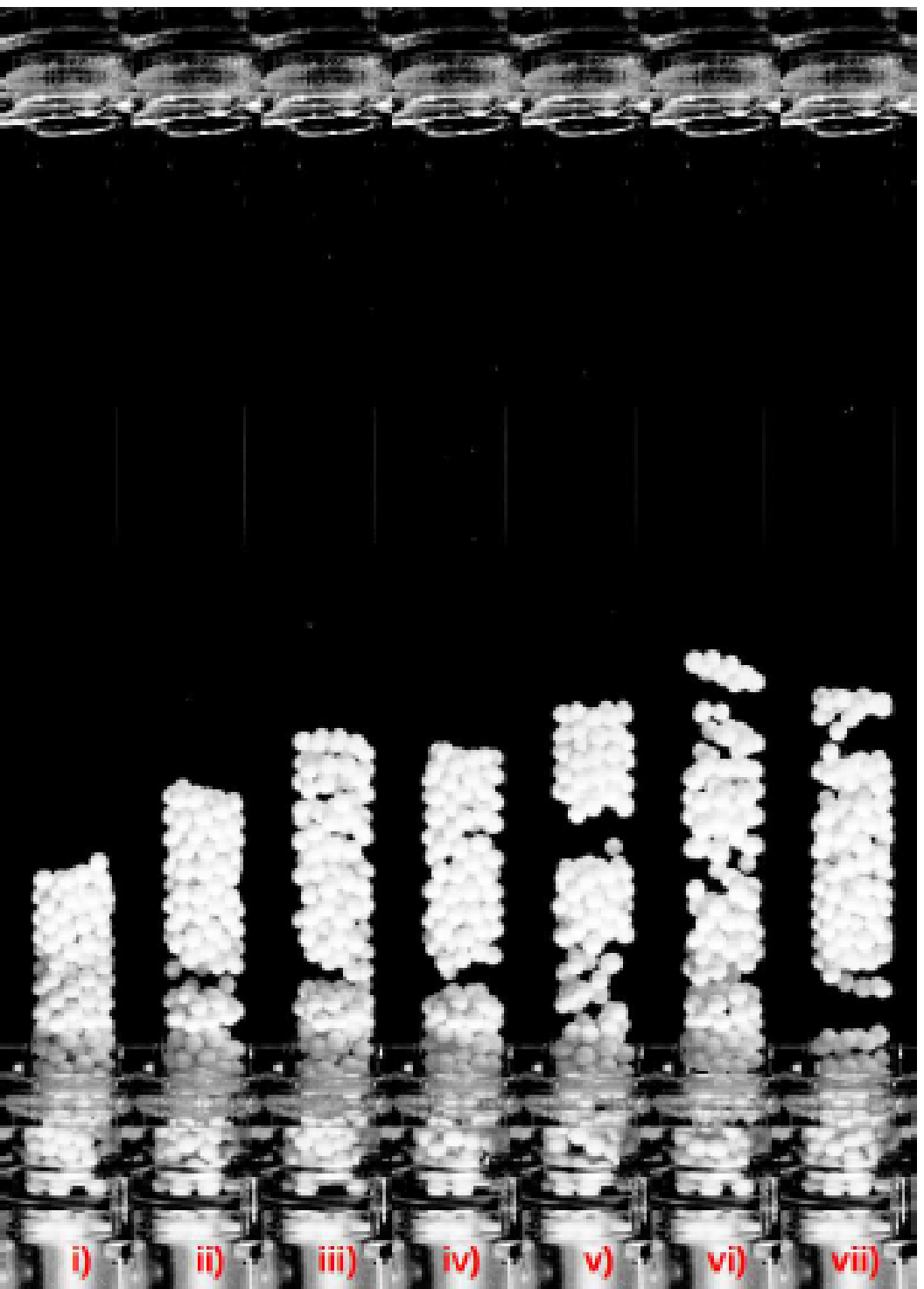}\\
 			(b)
 		\end{tabular}
 	\end{minipage}
	\hfill
	\caption{Instantaneous snapshots of particles positions: (a) $N$ = 250 and $\overline{U}$ = 0.137 m/s; (b) $N$ = 250 and $\overline{U}$ = 0.164 m/s. The corresponding times are: (i) 0 s; (ii) 1 s; (iii) 1.5 s; (iv) 2 s; (v) 2.5 s; (vi) 3 s; (vii) 3.5 s.}
	\label{fig:experimentos_grains250}
 \end{figure}

\begin{figure}[ht]
 	\begin{minipage}{0.5\linewidth}
 		\begin{tabular}{c}
 			\includegraphics[width=0.90\linewidth]{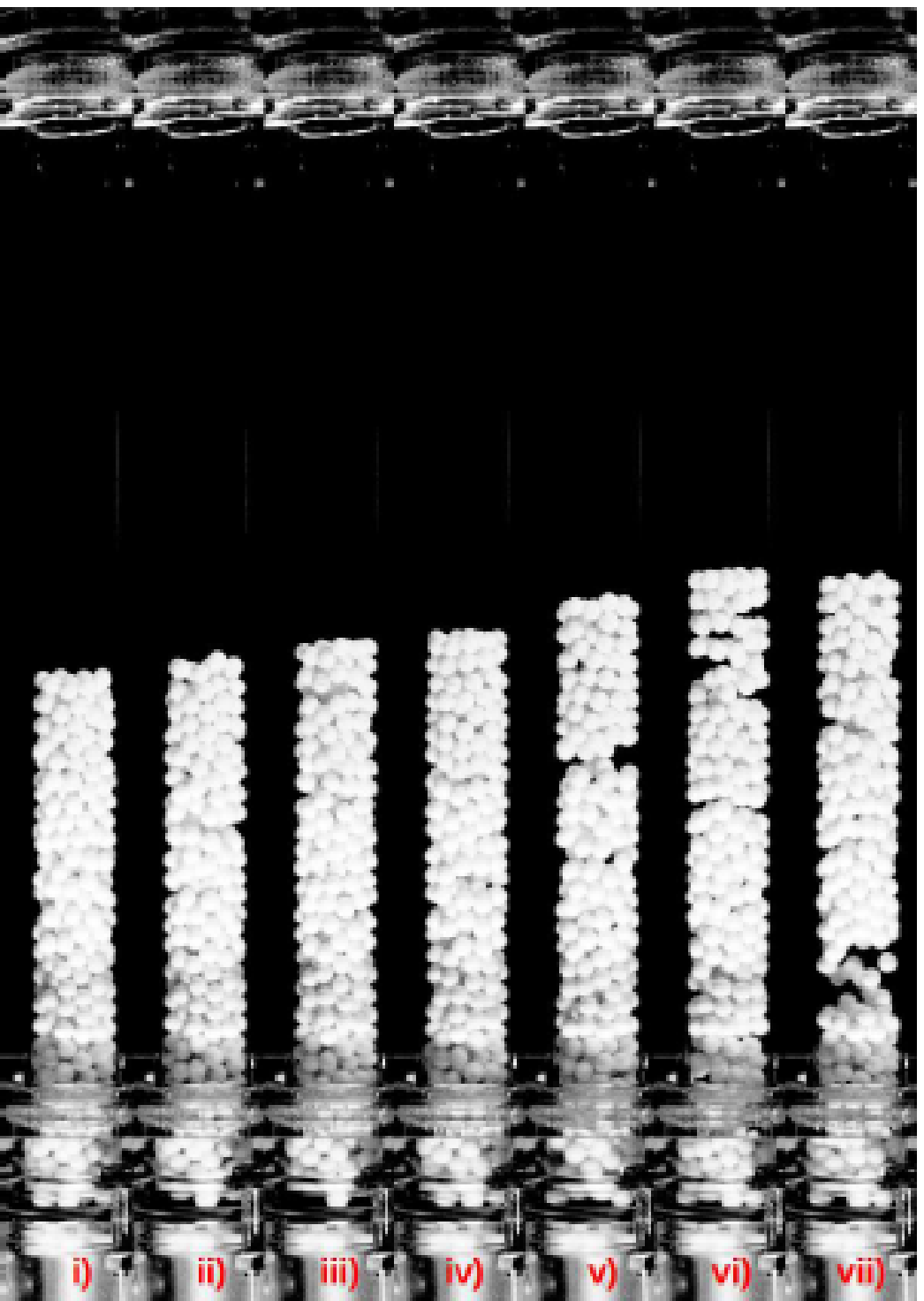}\\
 			(a)
 		\end{tabular}
 	\end{minipage}
 	\hfill
 	\begin{minipage}{0.5\linewidth}
 		\begin{tabular}{c}
 			\includegraphics[width=0.90\linewidth]{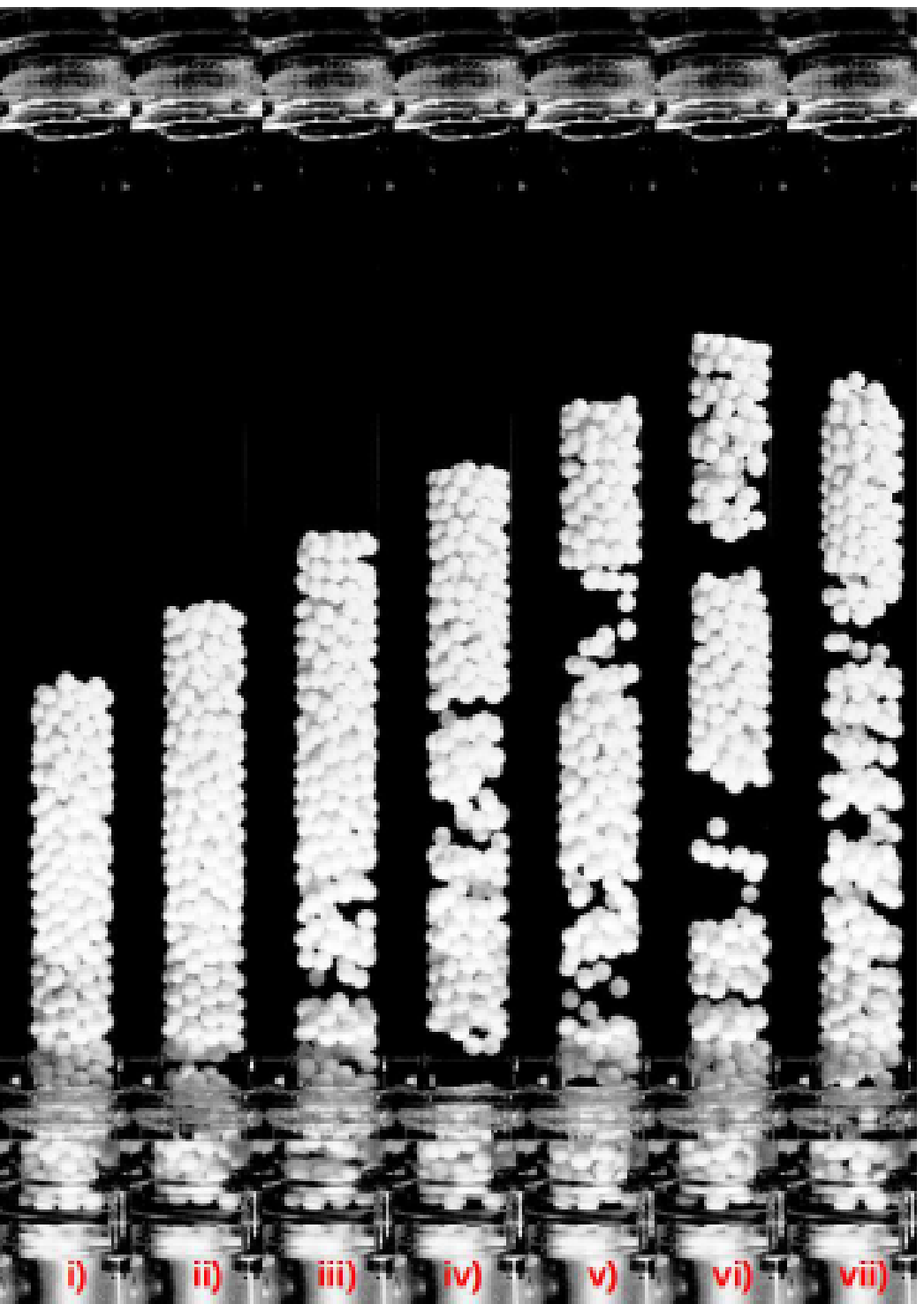}\\
 			(b)
 		\end{tabular}
 	\end{minipage}
	\hfill
	\caption{Instantaneous snapshots of particles positions: (a) $N$ = 400 and $\overline{U}$ = 0.137 m/s; (b) $N$ = 400 and $\overline{U}$ = 0.164 m/s. The corresponding times are: (i) 0 s; (ii) 1 s; (iii) 1.5 s; (iv) 2 s; (v) 2.5 s; (vi) 3 s; (vii) 3.5 s.}
	\label{fig:experimentos_grains400}
 \end{figure}

\begin{figure}[ht]
 	\begin{minipage}{0.5\linewidth}
 		\begin{tabular}{c}
 			\includegraphics[width=0.90\linewidth]{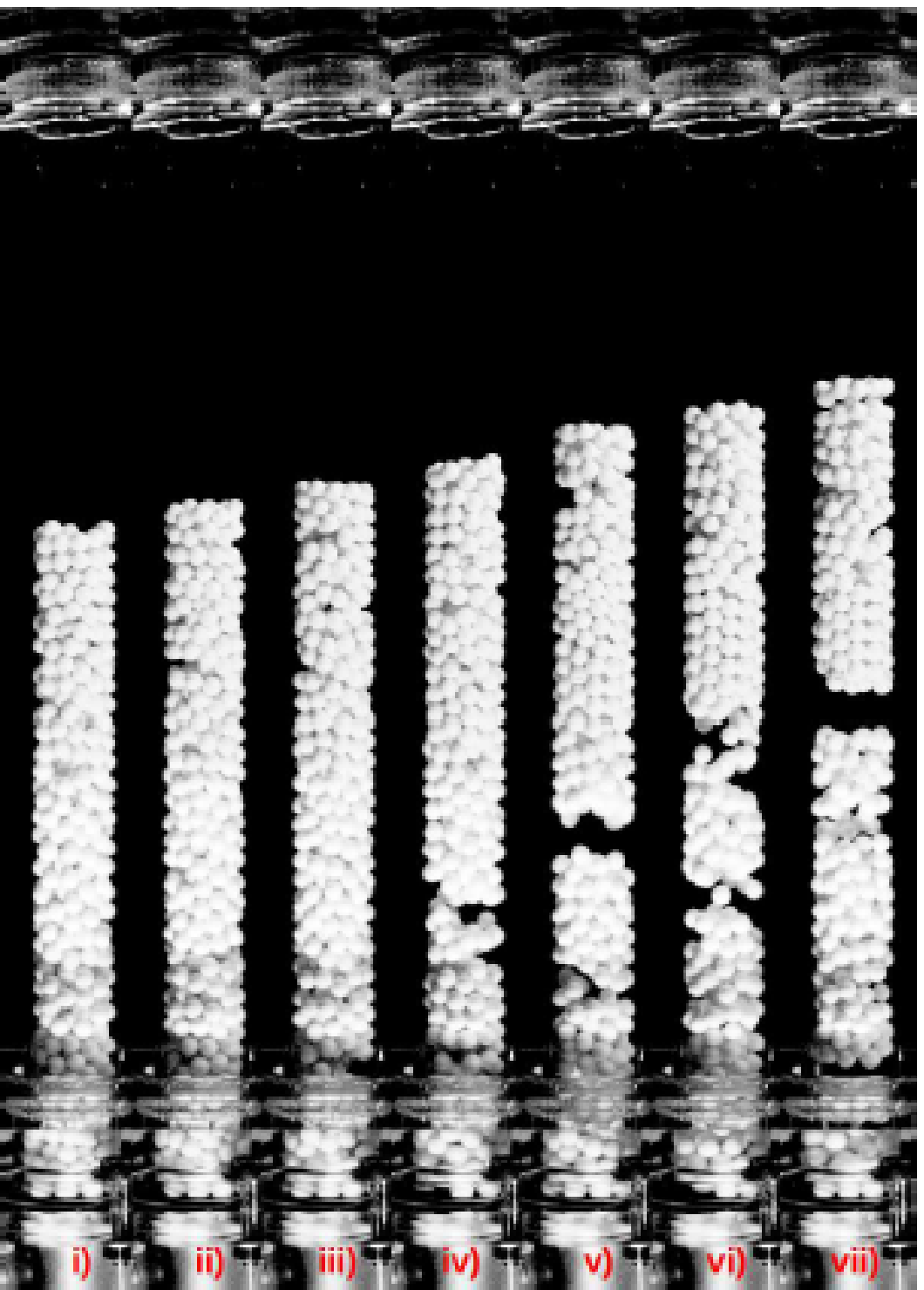}\\
 			(a)
 		\end{tabular}
 	\end{minipage}
 	\hfill
 	\begin{minipage}{0.5\linewidth}
 		\begin{tabular}{c}
 			\includegraphics[width=0.90\linewidth]{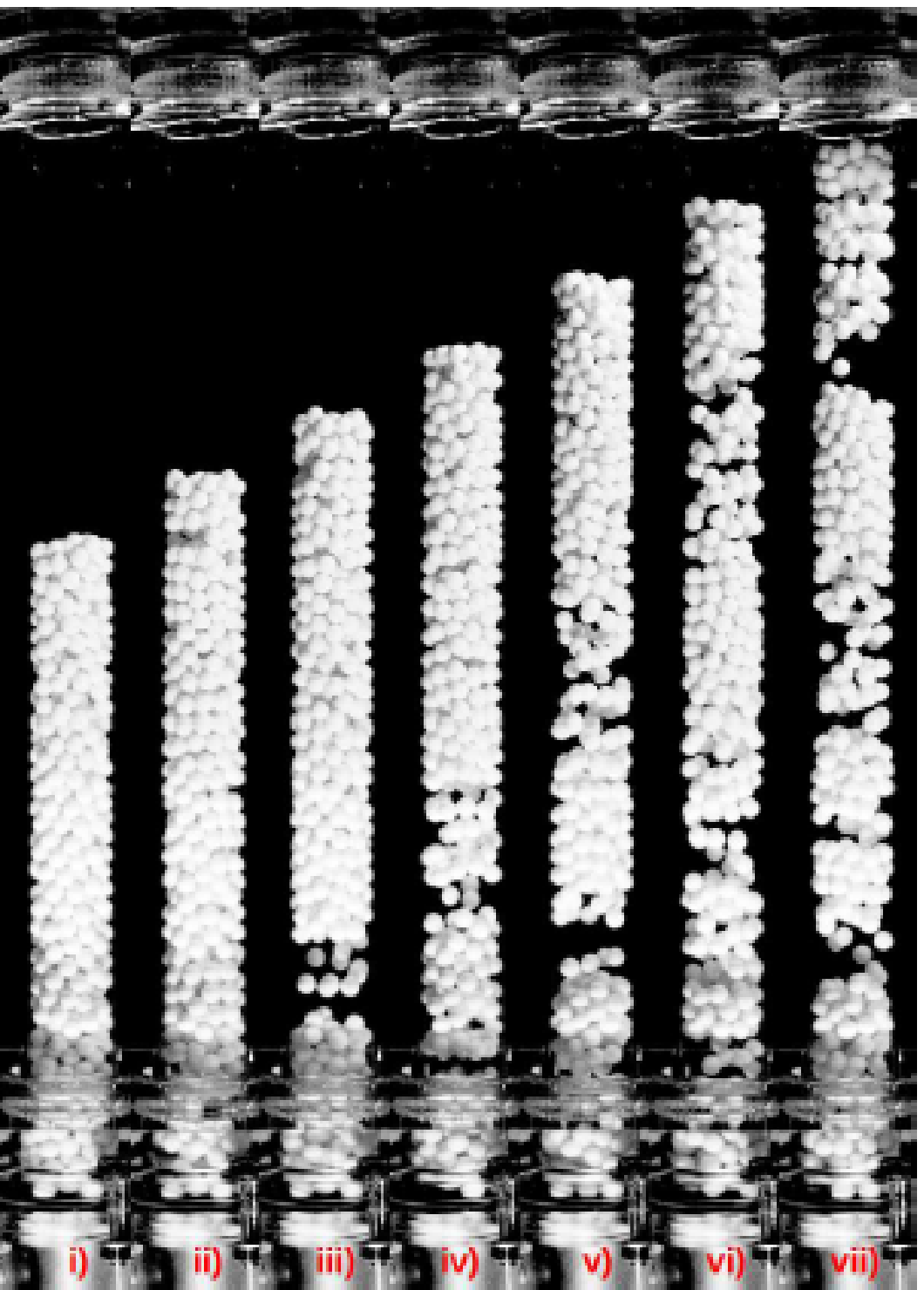}\\
 			(b)
 		\end{tabular}
 	\end{minipage}
	\hfill
	\caption{Instantaneous snapshots of particles positions: (a) $N$ = 500 and $\overline{U}$ = 0.137 m/s; (b) $N$ = 500 and $\overline{U}$ = 0.164 m/s. The corresponding times are: (i) 0 s; (ii) 1 s; (iii) 1.5 s; (iv) 2 s; (v) 2.5 s; (vi) 3 s; (vii) 3.5 s.}
	\label{fig:experimentos_grains500}
 \end{figure}

Figs. \ref{fig:experimentos_grains250}, \ref{fig:experimentos_grains400} and \ref{fig:experimentos_grains500} present some frames obtained with the high-speed camera for the 250, 400 and 500 particles beds, respectively, for both flow rates. The corresponding times are in the caption of figures. From Figs. \ref{fig:experimentos_grains250} to \ref{fig:experimentos_grains500}, we can observe the plugs and bubbles in the bed. In order to give the reader a better idea of the bed dynamics, a movie from one experiment and an animation from one numerical simulation are attached to this paper as Supplementary Material \cite{Supplemental}.

The upward propagation of the void regions made the top of the bed oscillate between minimum and maximum values. Because of this oscillation, we computed the average height of the fluidized bed $h_{avg}$ as the average between minimum and maximum values. In addition, we computed the upward, $C_{up}$, and downward, $C_{down}$, celerities of the top as the derivative of the measured positions during its rise and descent, respectively.

We identified and followed each granular plug in the high-speed movies by using numerical scripts whose description can be found in \cite{Alvarez}. We present next the length scales and the celerities obtained from the experiments. Tab. \ref{table:table2} presents the superficial velocity $\overline{U}$, number of grains $N$, initial bed height $h_{mf}$, length scale of plugs $\lambda$, standard deviation of length scale $\sigma_\lambda$, length scale of plugs normalized by the grain diameter $\lambda /d$, standard deviation of length scale normalized by the grain diameter $\sigma_\lambda /d$, bed upward celerity $C_{up}$, standard deviation of upward celerities $\sigma_{C,up}$, bed downward celerity $C_{down}$, standard deviation of downward celerities $\sigma_{C,down}$, and average height of the fluidized bed $h_{avg}$.

\begin{table}[ht]
\caption{Superficial velocity $\overline{U}$, Number of grains $N$, initial bed height $h_{mf}$, length scale of plugs $\lambda$, standard deviation of length scale $\sigma_\lambda$, length scale of plugs normalized by the grain diameter $\lambda /d$, standard deviation of length scale normalized by the grain diameter $\sigma_\lambda /d$, bed upward celerity $C_{up}$, standard deviation of upward celerities $\sigma_{C,up}$, bed downward celerity $C_{down}$, standard deviation of downward celerities $\sigma_{C,down}$, and average height of the fluidized bed $h_{avg}$, obtained from experiments.}
\label{table:table2}
\centering
\begin{tabular}{c c c c c c c c}
\hline \hline

case & $\cdots$ & (a) & (b) & (c) & (d) & (e) & (f)\\
$\overline{U}$ & m/s & 0.137 & 0.137 & 0.137 & 0.164 & 0.164 & 0.164\\
$N$ & $\cdots$ & 250 & 400 & 500 & 250 & 400 & 500\\
$h_{mf}$ & m & 0.117 & 0.181 & 0.225 & 0.117 & 0.181 & 0.225\\
$\lambda$ & m & 0.075 & 0.082 & 0.072 & 0.040 & 0.041 & 0.042\\
$\sigma_\lambda$ & m & 0.027 & 0.047 & 0.046 & 0.026 & 0.028 & 0.030\\
$\lambda/d$ & $\cdots$ & 12.4 & 13.7 & 11.9 & 6.6 & 6.8 & 7.1\\
$\sigma_\lambda /d$ & $\cdots$ & 4.4 & 7.9 & 7.6 & 4.3 & 4.6 & 5.0\\
$C_{up}$ & m/s & 0.012 & 0.017 & 0.018 & 0.029 & 0.035 & 0.060\\
$\sigma_{C,up}$ & m/s & 0.007 & 0.009 & 0.008 & 0.016 & 0.017 & 0.061\\
$C_{down}$ & m/s & -0.019 & -0.029 & -0.040 & -0.049 & -0.057 & -0.114\\
$\sigma_{C,down}$ & m/s & 0.017 & 0.028 & 0.049 & 0.038 & 0.046 & 0.099\\
$h_{avg}$ & m & 0.142 & 0.226 & 0.287 & 0.177 & 0.287 & 0.354\\

\hline 
\end{tabular} 
\end{table}

Within the experimental conditions, the lengths of plugs show strong variations with the water flow and are independent of the initial height of the bed (or of the number of beads), despite the small variation in water superficial velocities (of only 20 \%) when compared to the variation in the number of particles (of 100 \%). The lengths of plugs were around 12$d$ for $\overline{U}$ = 0.137 m/s and 7$d$ for $\overline{U}$ = 0.164 m/s. This represents a variation of 100 \% in $\lambda$ for variations of 20 \% in $\overline{U}$. Standard deviations of plug lengths are large. The reason for that is the discrete nature of plugs, which consist of solid particles that are large with respect to the plugs: the grain diameter is around 10 \% of the plug length and 25 \% of the tube diameter.

Bed celerities present variations with both the initial bed height and flow conditions. From $N$ = 250 to $N$ = 500, $C_{up}$ varies from 0.012 to 0.018 m/s for $\overline{U}$ = 0.137 m/s and from 0.029 to 0.069 m/s for $\overline{U}$ = 0.164 m/s, while $C_{down}$ varies from -0.019 to -0.040 m/s for $\overline{U}$ = 0.137 m/s and from -0.049 to -0.114 m/s for $\overline{U}$ = 0.164 m/s. This represents variations of 130 \% in $C_{up}$ and $C_{down}$ for variations of 100 \% in $N$, and of 230 \% in $C_{up}$ and 190 \% in $C_{down}$ for variations of 20 \% in $\overline{U}$. As for $\lambda$, standard deviations of the celerities are large due to the relatively large size of solid particles.

\subsection{Numerical Simulations}

The numerical simulations were performed with the same parameters used in the experiments; therefore, comparisons can be made directly with the experimental results. The main advantage of the CFD-DEM simulations when compared to experiments is that some parameters difficult to be obtained from experiments are readily available in simulations. Some examples are the displacement of individual grains within the bed, the local fluid velocity, the local void fraction, and the network of contact forces. We present next the results concerning the granular patterns of the bed.

\begin{figure}[ht]
 	\begin{minipage}{0.5\linewidth}
 		\begin{tabular}{c}
 			\includegraphics[width=0.90\linewidth]{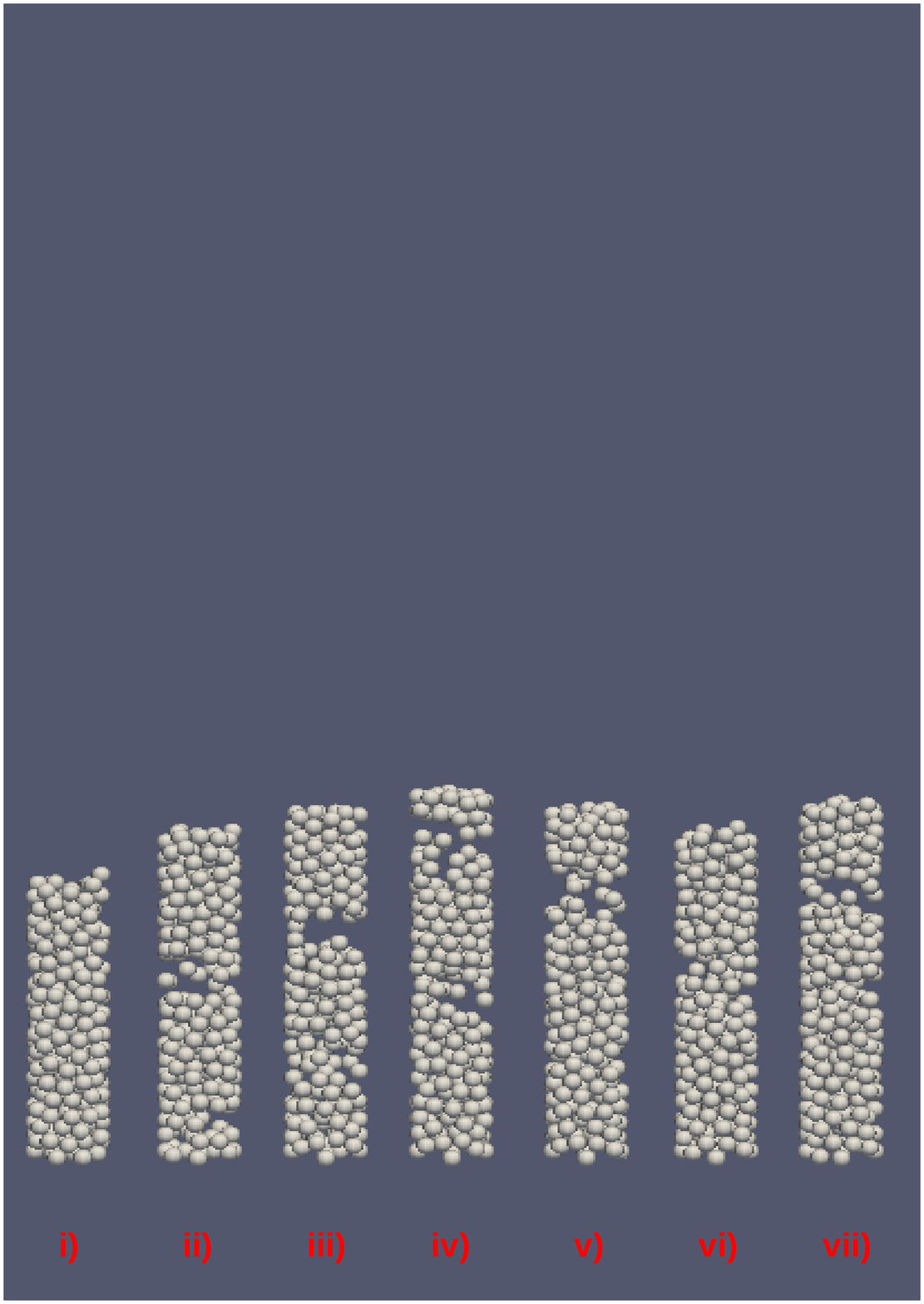}\\
 			(a)
 		\end{tabular}
 	\end{minipage}
 	\hfill
 	\begin{minipage}{0.5\linewidth}
 		\begin{tabular}{c}
 			\includegraphics[width=0.90\linewidth]{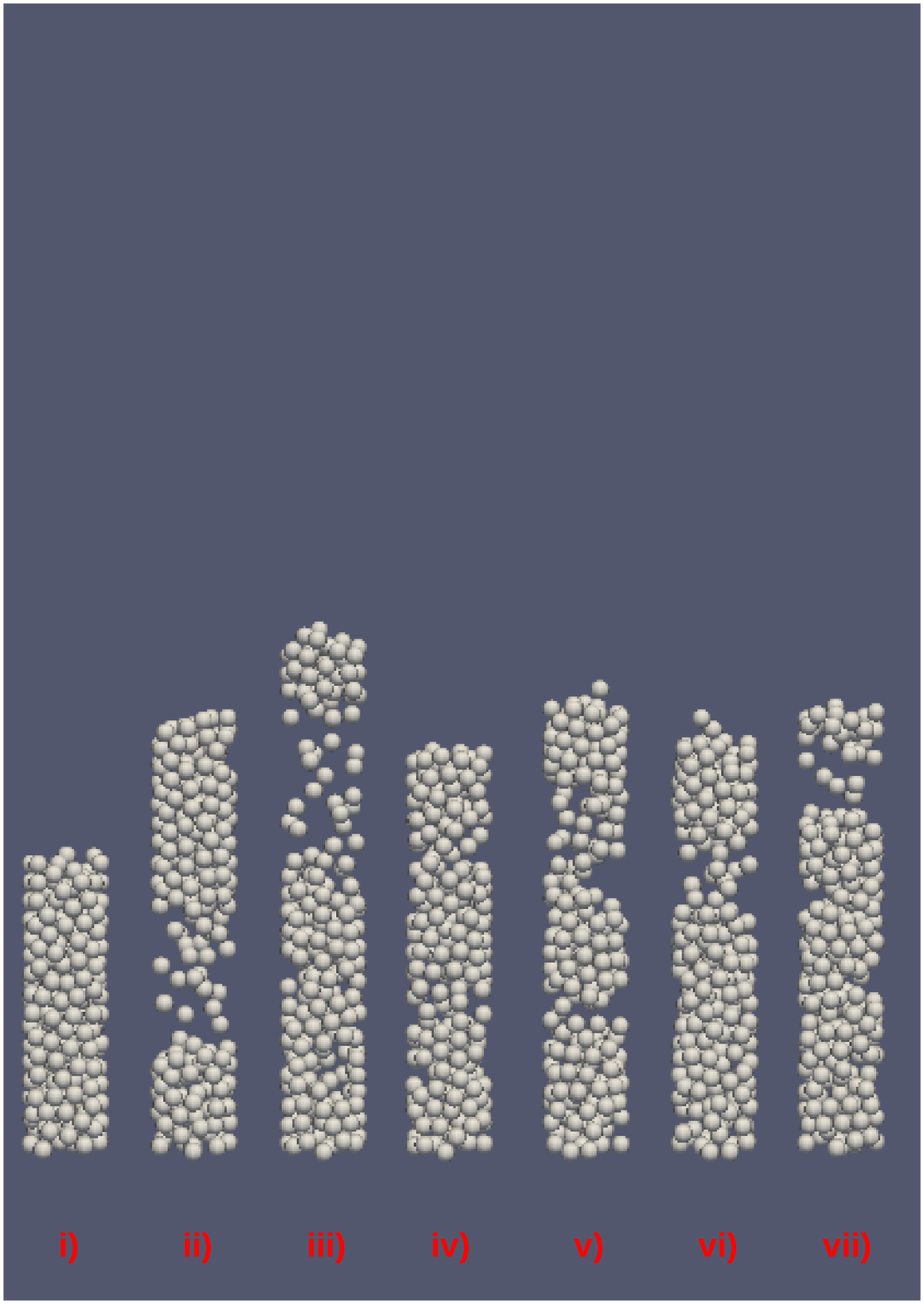}\\
 			(b)
 		\end{tabular}
 	\end{minipage}
	\hfill
	\caption{Instantaneous snapshots of particles positions: (a) $N$ = 250 and $\overline{U}$ = 0.137 m/s; (b) $N$ = 250 and $\overline{U}$ = 0.164 m/s. The corresponding times are: (i) 0 s; (ii) 1 s; (iii) 1.5 s; (iv) 2 s; (v) 2.5 s; (vi) 3 s; (vii) 3.5 s.}
	\label{fig:num_simulations_grains250}
 \end{figure}

\begin{figure}[ht]
 	\begin{minipage}{0.5\linewidth}
 		\begin{tabular}{c}
 			\includegraphics[width=0.90\linewidth]{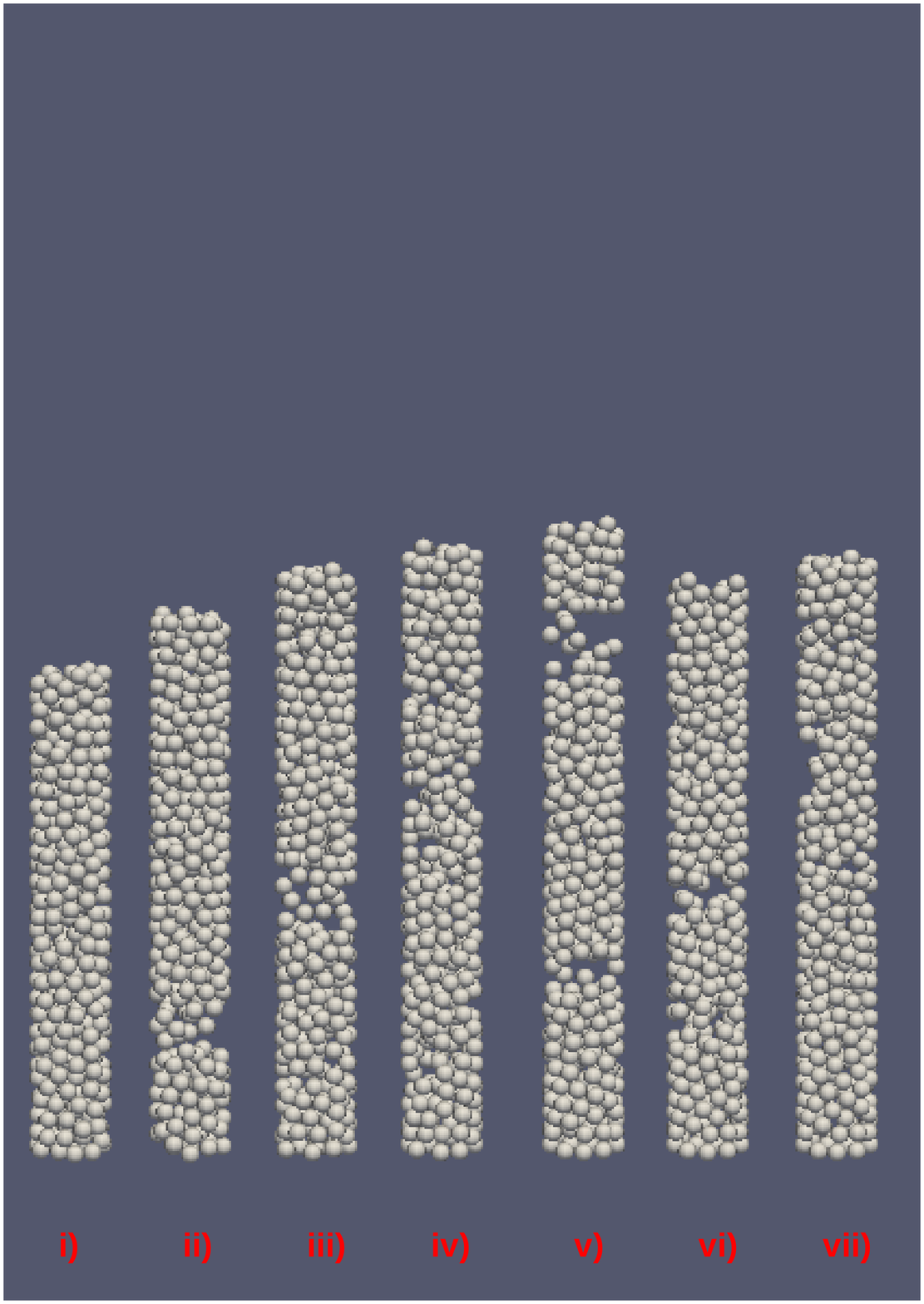}\\
 			(a)
 		\end{tabular}
 	\end{minipage}
 	\hfill
 	\begin{minipage}{0.5\linewidth}
 		\begin{tabular}{c}
 			\includegraphics[width=0.90\linewidth]{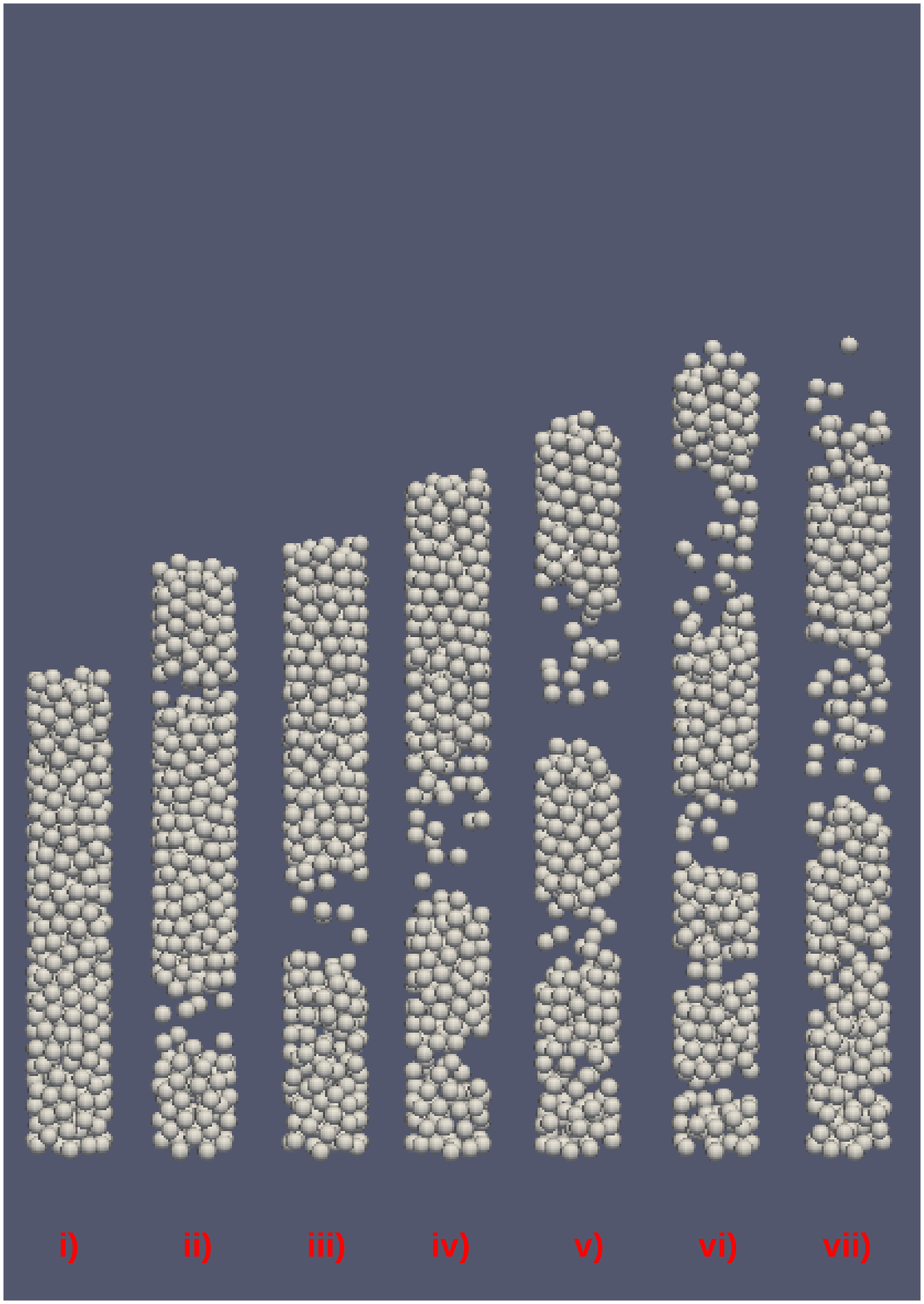}\\
 			(b)
 		\end{tabular}
 	\end{minipage}
	\hfill
	\caption{Instantaneous snapshots of particles positions: (a) $N$ = 400 and $\overline{U}$ = 0.137 m/s; (b) $N$ = 400 and $\overline{U}$ = 0.164 m/s. The corresponding times are: (i) 0 s; (ii) 1 s; (iii) 1.5 s; (iv) 2 s; (v) 2.5 s; (vi) 3 s; (vii) 3.5 s.}
	\label{fig:num_simulations_grains400}
 \end{figure}

\begin{figure}[ht]
 	\begin{minipage}{0.5\linewidth}
 		\begin{tabular}{c}
 			\includegraphics[width=0.90\linewidth]{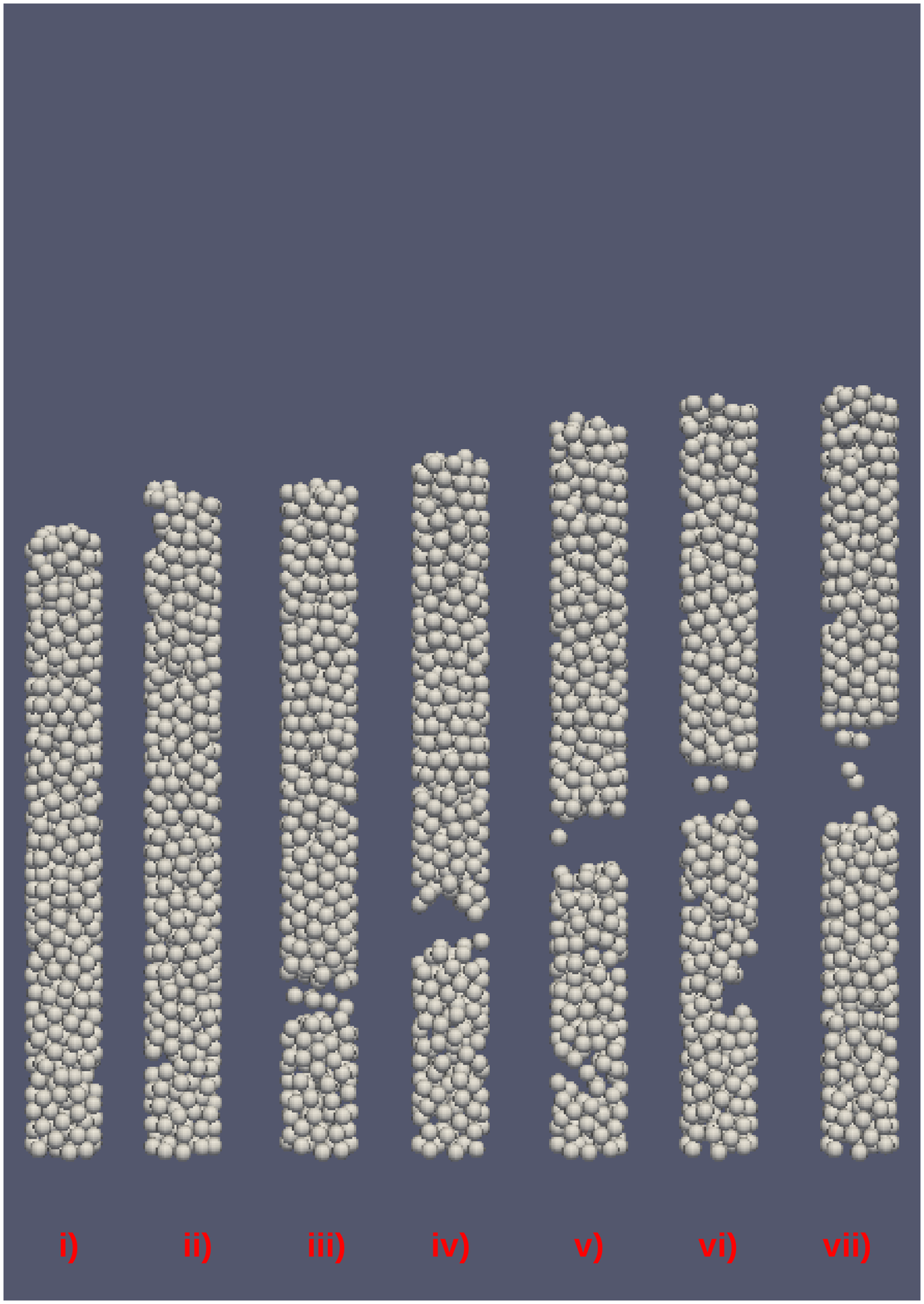}\\
 			(a)
 		\end{tabular}
 	\end{minipage}
 	\hfill
 	\begin{minipage}{0.5\linewidth}
 		\begin{tabular}{c}
 			\includegraphics[width=0.90\linewidth]{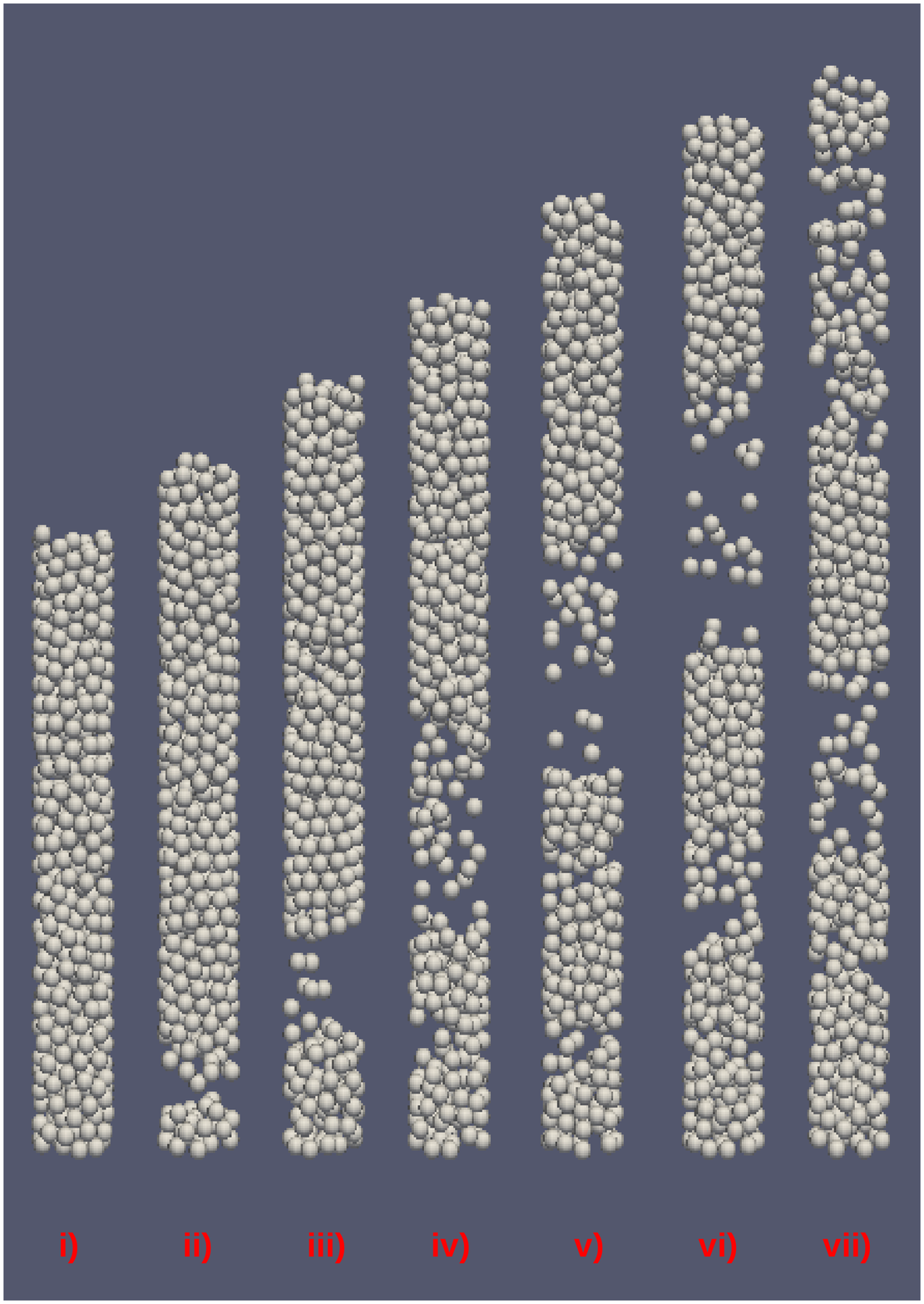}\\
 			(b)
 		\end{tabular}
 	\end{minipage}
	\hfill
	\caption{Instantaneous snapshots of particles positions: (a) $N$ = 500 and $\overline{U}$ = 0.137 m/s; (b) $N$ = 500 and $\overline{U}$ = 0.164 m/s. The corresponding times are: (i) 0 s; (ii) 1 s; (iii) 1.5 s; (iv) 2 s; (v) 2.5 s; (vi) 3 s; (vii) 3.5 s.}
	\label{fig:num_simulations_grains500}
 \end{figure}

Figs. \ref{fig:num_simulations_grains250}, \ref{fig:num_simulations_grains400} and \ref{fig:num_simulations_grains500} present instantaneous snapshots of particles positions for the 250, 400 and 500 particles beds, respectively, for both flow rates. The corresponding times are in the caption of figures. Comparing Figs. \ref{fig:num_simulations_grains250} to \ref{fig:num_simulations_grains500} with Figs. \ref{fig:experimentos_grains250} to \ref{fig:experimentos_grains500}, we observe that the pattern formation is well captured by the numerical simulations.

The granular patterns filmed in experiments (Figs. \ref{fig:experimentos_grains250} to \ref{fig:experimentos_grains500}) and formed in numerical simulations (Figs. \ref{fig:num_simulations_grains250} to \ref{fig:num_simulations_grains500}) have a strong influence on the fluid flow due to the movement of individual grains and the presence of granular plugs. These small and large granular structures interact with the fluid flow creating specific patterns for the local fluid velocities and local void fractions. Supplementary Material \cite{Supplemental2} attached to this paper shows the instantaneous snapshots of water flow velocities and local void fractions for the beds consisting of 400 and 500 particles. Those values of local void fraction and water velocity are difficult to obtain from experiments and are closely related to the wavy patterns of the bed. The use of the local void fraction is another option to measure the length of granular plugs; however, in order to be consistent with the image treatment from experimental measurements, we computed the plug lengths and bed celerities from the identification of solid particles.

We identified and followed each granular plug by using numerical scripts written in the course of this work. We present next the length scales and celerities obtained from the numerical simulations. Tab. \ref{table:table3} presents the superficial velocity $\overline{U}$, number of grains $N$, initial bed height $h_{mf}$, length scale of plugs $\lambda$, standard deviation of length scale $\sigma_\lambda$, length scale of plugs normalized by the grain diameter $\lambda /d$, standard deviation of length scale normalized by the grain diameter $\sigma_\lambda /d$, bed upward celerity $C_{up}$, standard deviation of upward celerities $\sigma_{C,up}$, bed downward celerity $C_{down}$, standard deviation of downward celerities $\sigma_{C,down}$, and average height of the fluidized bed $h_{avg}$. 

\begin{table}[ht]
\caption{Superficial velocity $\overline{U}$, Number of grains $N$, initial bed height $h_{mf}$, length scale of plugs $\lambda$, standard deviation of length scale $\sigma_\lambda$, length scale of plugs normalized by the grain diameter $\lambda /d$, standard deviation of length scale normalized by the grain diameter $\sigma_\lambda /d$, bed upward celerity $C_{up}$, standard deviation of upward celerities $\sigma_{C,up}$, bed downward celerity $C_{down}$, standard deviation of downward celerities $\sigma_{C,down}$, and average height of the fluidized bed $h_{avg}$, obtained from numerical simulations.}
\label{table:table3}
\centering
\begin{tabular}{c c c c c c c c}
\hline \hline
case & $\cdots$ & (a) & (b) & (c) & (d) & (e) & (f)\\
$\overline{U}$ & m/s & 0.137 & 0.137 & 0.137 & 0.164 & 0.164 & 0.164\\
$N$ & $\cdots$ & 250 & 400 & 500 & 250 & 400 & 500\\
$h_{mf}$ & m & 0.111 & 0.177 & 0.223 & 0.111 & 0.177 & 0.223\\
$\lambda$ & m & 0.091 & 0.102 & 0.095 & 0.052 & 0.055 & 0.056\\
$\sigma_\lambda$ & m & 0.037 & 0.055 & 0.051 & 0.031 & 0.026 & 0.030\\
$\lambda/d$ & $\cdots$ & 15.1 & 17.0 & 15.8 & 8.6 & 9.2 & 9.3\\
$\sigma_\lambda /d$ & $\cdots$ & 6.1 & 9.2 & 8.5 & 5.2 & 4.3 & 5.1\\
$C_{up}$ & m/s & 0.013 & 0.015 & 0.018 & 0.033 & 0.032 & 0.049\\
$\sigma_{C,up}$ & m/s & 0.007 & 0.009 & 0.008 & 0.022 & 0.015 & 0.050\\
$C_{down}$ & m/s & -0.021 & -0.022 & -0.039 & -0.047 & -0.063 & -0.115\\
$\sigma_{C,down}$ & m/s & 0.020 & 0.022 & 0.055 & 0.040 & 0.054 & 0.084\\
$h_{avg}$ & m & 0.138 & 0.217 & 0.274 & 0.170 & 0.266 & 0.340\\
\hline 
\end{tabular} 
\end{table}

As in experiments, the lengths of plugs show strong variations with the water flow and are independent of the initial height of the bed. The lengths of plugs were around 16$d$ for $\overline{U}$ = 0.137 m/s and 9$d$ for $\overline{U}$ = 0.164 m/s. Standard deviations of plug lengths are large due to the discrete nature of plugs. Bed celerities present variations with both the initial bed height and flow conditions. From $N$ = 250 to $N$ = 500, $C_{up}$ varies from 0.013 to 0.018 m/s for $\overline{U}$ = 0.137 m/s and from 0.033 to 0.049 m/s for $\overline{U}$ = 0.164 m/s, while $C_{down}$ varies from -0.021 to -0.039 m/s for $\overline{U}$ = 0.137 m/s and from -0.047 to -0.115 m/s for $\overline{U}$ = 0.164 m/s. The values of celerities are close to the experimentally measured ones. Standard deviations of celerities are large due to the relatively large size of solid particles. The behavior and tendencies of lengths and celerities are similar to the experimental observations.

By comparing the results of simulations with experimental results, we note differences of the order of 10 \% between them with the exception of plug lengths, which present differences of the order of 20 to 30 \%. Considering that the measured structures consist of discrete particles which are large with respect to the structures (the grain diameter is around 10 \% of the plug length and 25 \% of the tube diameter), these differences are within the expected errors: the arrival (or departure) of an individual particle on a plug may represent a difference of 10 \% in the measured length (and celerity). In the particular case of plug lengths, the relatively higher differences between simulations and experiments are caused by the differences in post-treatment. While in experiments we obtained images of front views of the tube interior, showing only particles in contact with the tube inner wall, in simulations we obtained the positions of each individual particle. The post-treatment was different in each case, leading to differences in the computed results that, added to the discrete nature of plugs, increased differences between experiments and simulations.

For the top of the bed, the arrival and departure of new grains were rare and it oscillated in a quasi one-dimensional way, with the top of grains forming a plane in the tube cross section \cite{Supplemental}. In this case, smaller differences between experiments and simulations are expected for the values of $h_{avg}$, which was computed as the average between minimum and maximum values of the top of the bed. Indeed, differences in $h_{avg}$ are between 3 and 7 \%, corroborating that differences were increased by post-treatment procedures. Taking into account the discrete nature of granular media, as well as the differences in processing experimental and numerical data, the values of the length scales and celerities obtained from simulations are in good agreement with experimental results.

Finally, we computed the network of contact forces for all the tested cases. The evolution of the chains of contact forces can determine the influence of the lateral walls in the formation of plugs. An animation showing the evolution of the network of contact forces from a numerical simulation is attached to this paper as Supplementary Material \cite{Supplemental}, together with the bed evolution. From this animation, we can identify the contact chains during the formation of granular plugs.

\begin{figure}[ht]
 	\begin{minipage}{0.5\linewidth}
 		\begin{tabular}{c}
 			\includegraphics[width=0.90\linewidth]{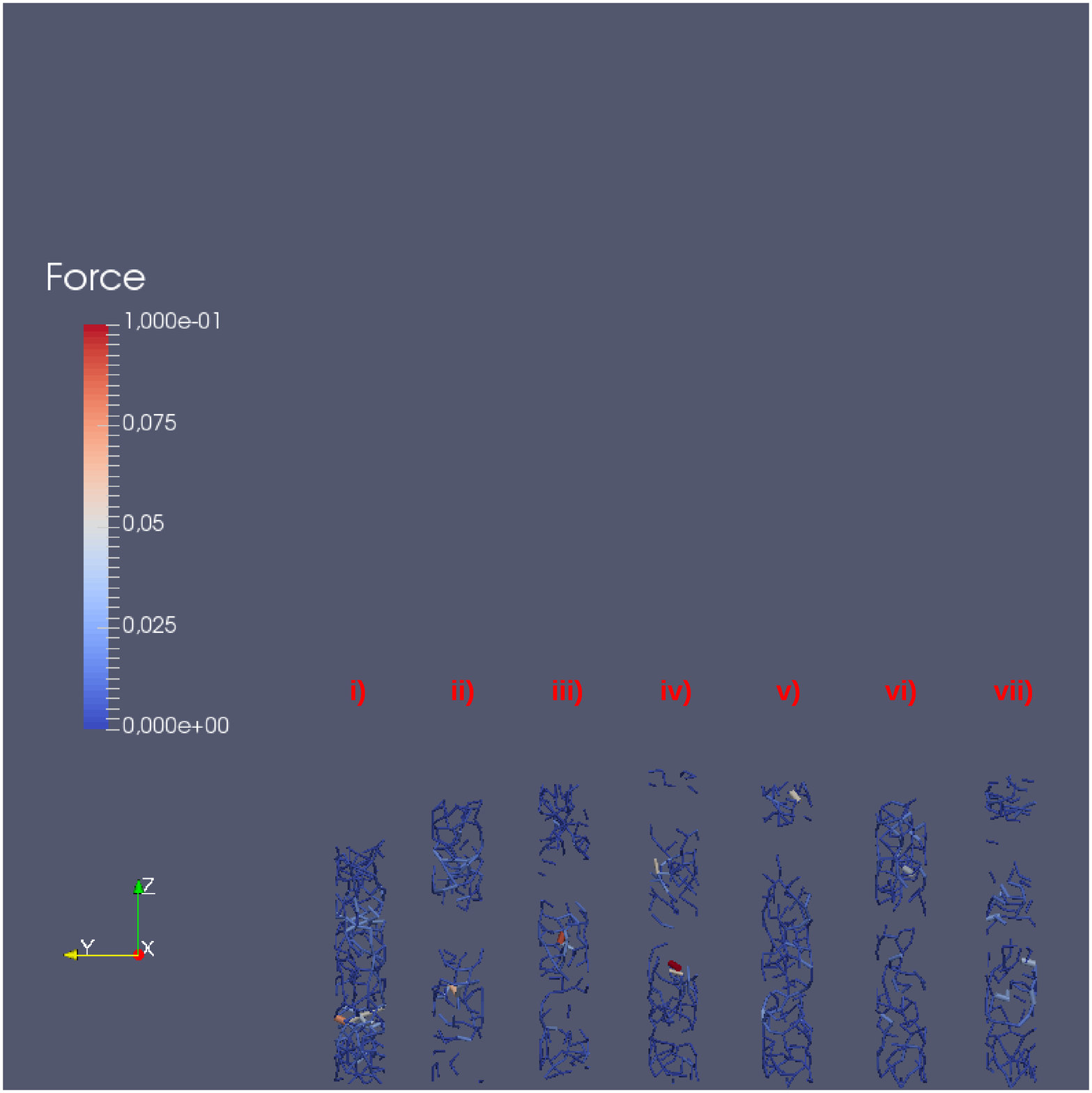}\\
 			(a)
 		\end{tabular}
 	\end{minipage}
 	\hfill
 	\begin{minipage}{0.5\linewidth}
 		\begin{tabular}{c}
 			\includegraphics[width=0.90\linewidth]{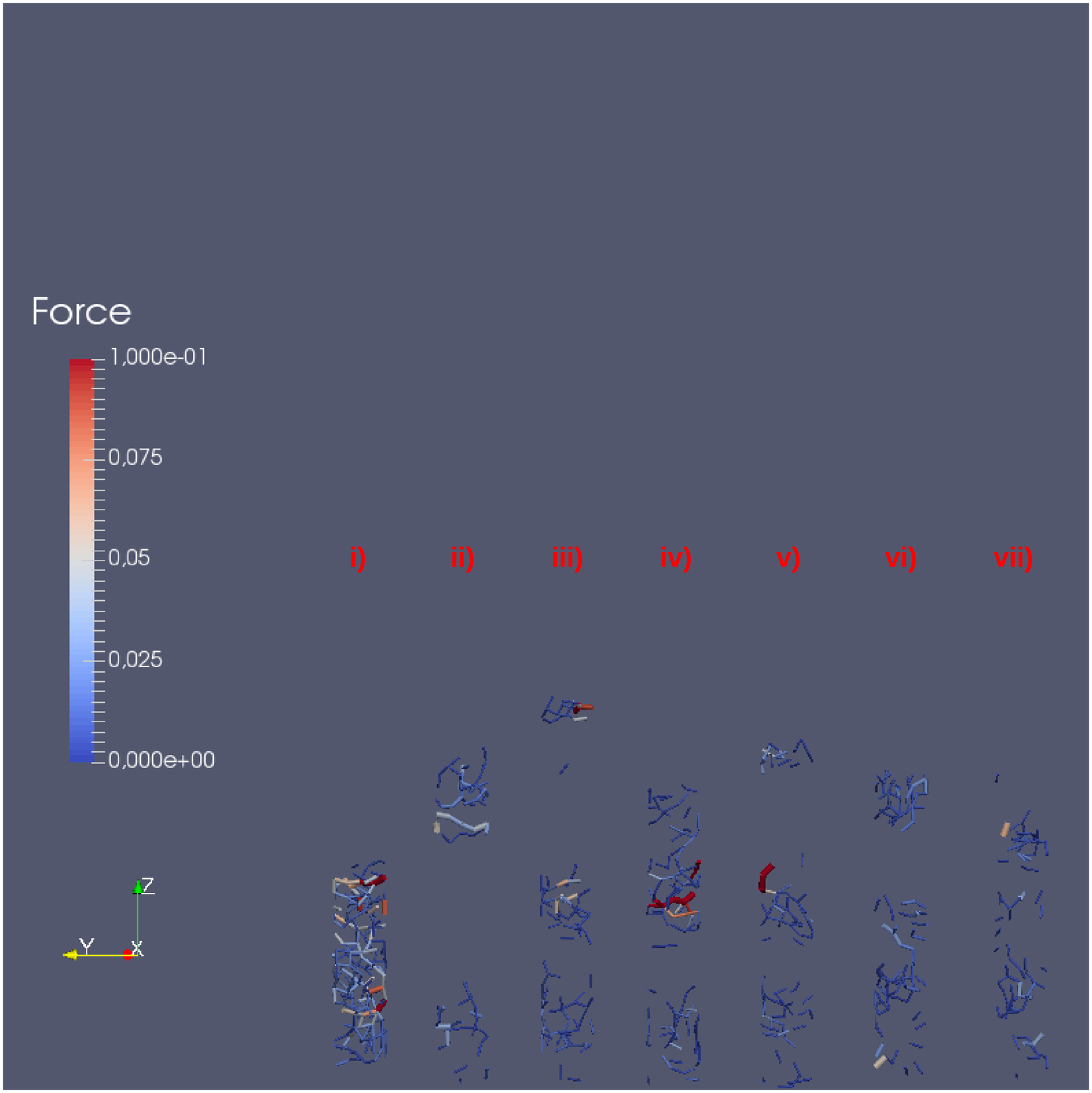}\\
 			(b)
 		\end{tabular}
 	\end{minipage}
	\hfill
	\caption{Instantaneous snapshots of the network of contact forces: (a) $N$ = 250 and $\overline{U}$ = 0.137 m/s; (b) $N$ = 250 and $\overline{U}$ = 0.164 m/s. The corresponding times are: (i) 0 s; (ii) 1 s; (iii) 1.5 s; (iv) 2 s; (v) 2.5 s; (vi) 3 s; (vii) 3.5 s. Values are in N.}
	\label{fig:num_contact_chains250}
 \end{figure}

\begin{figure}[ht]
 	 	\begin{minipage}{0.5\linewidth}
 		\begin{tabular}{c}
 			\includegraphics[width=0.90\linewidth]{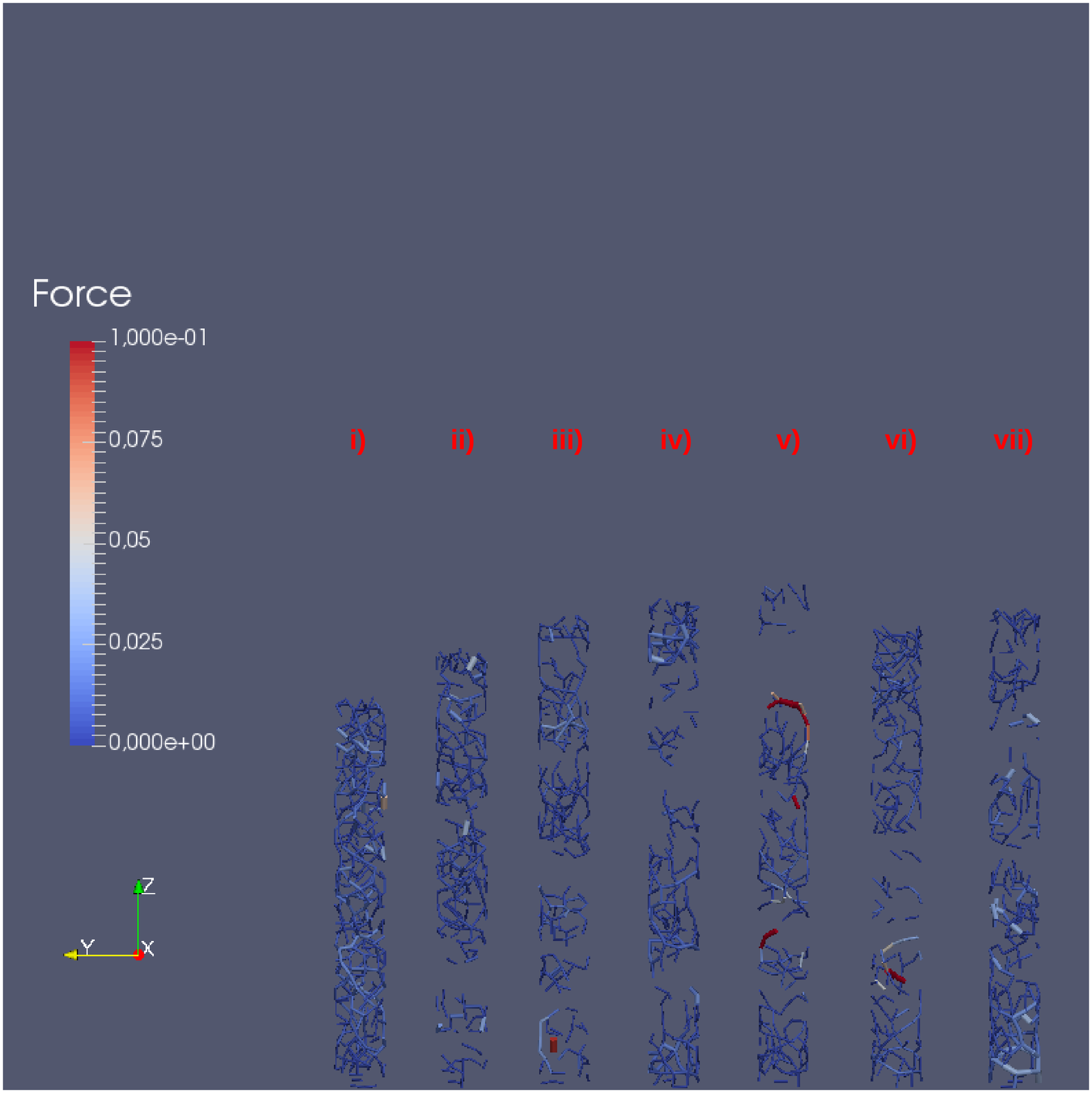}\\
 			(a)
 		\end{tabular}
 	\end{minipage}
 	\hfill
 	\begin{minipage}{0.5\linewidth}
 		\begin{tabular}{c}
 			\includegraphics[width=0.90\linewidth]{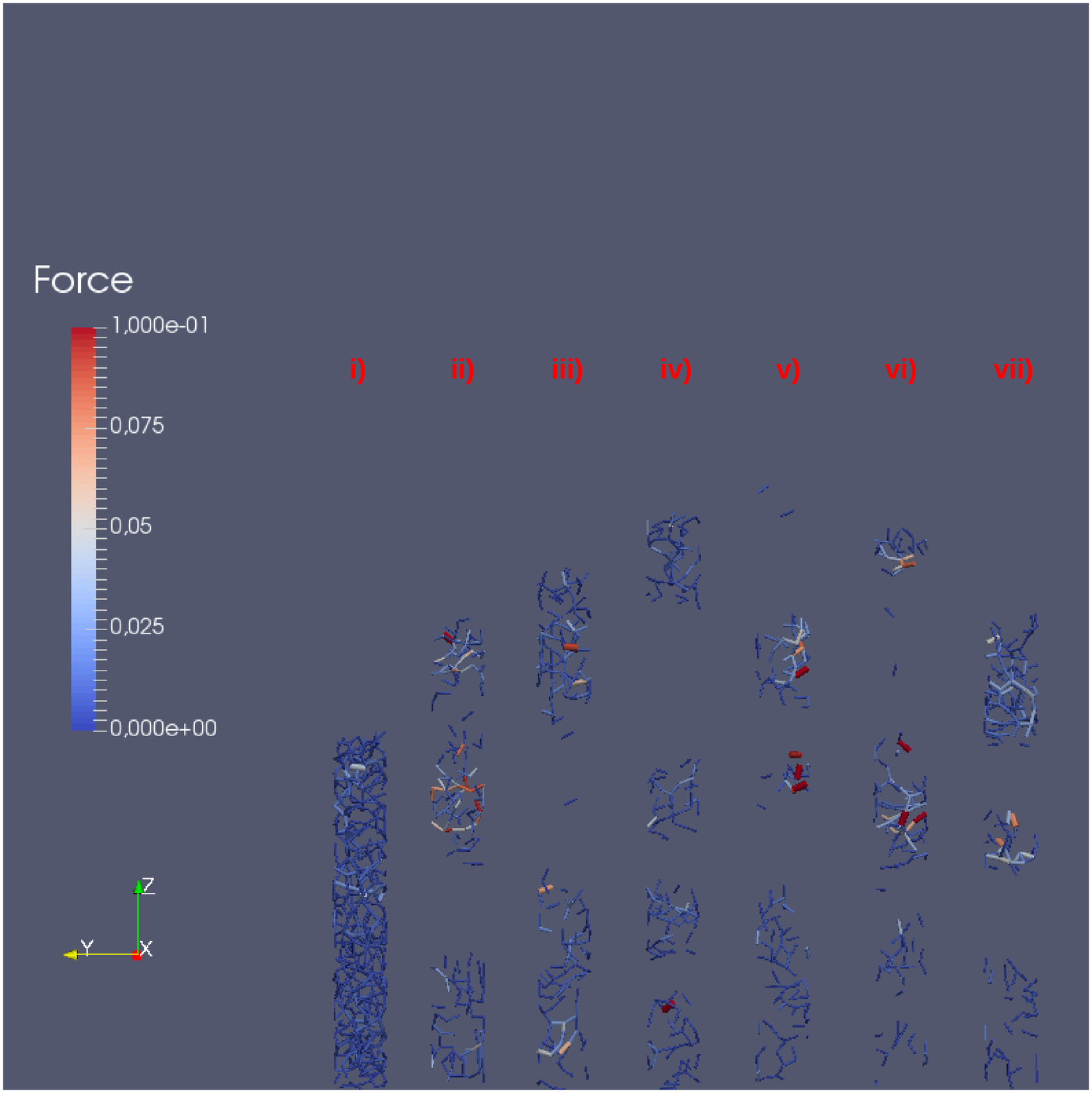}\\
 			(b)
 		\end{tabular}
 	\end{minipage}
	\hfill
	\caption{Instantaneous snapshots of the network of contact forces: (a) $N$ = 400 and $\overline{U}$ = 0.137 m/s; (b) $N$ = 400 and $\overline{U}$ = 0.164 m/s. The corresponding times are: (i) 0 s; (ii) 1 s; (iii) 1.5 s; (iv) 2 s; (v) 2.5 s; (vi) 3 s; (vii) 3.5 s. Values are in N.}
	\label{fig:num_contact_chains400}
 \end{figure}

\begin{figure}[ht]
 	 	\begin{minipage}{0.5\linewidth}
 		\begin{tabular}{c}
 			\includegraphics[width=0.90\linewidth]{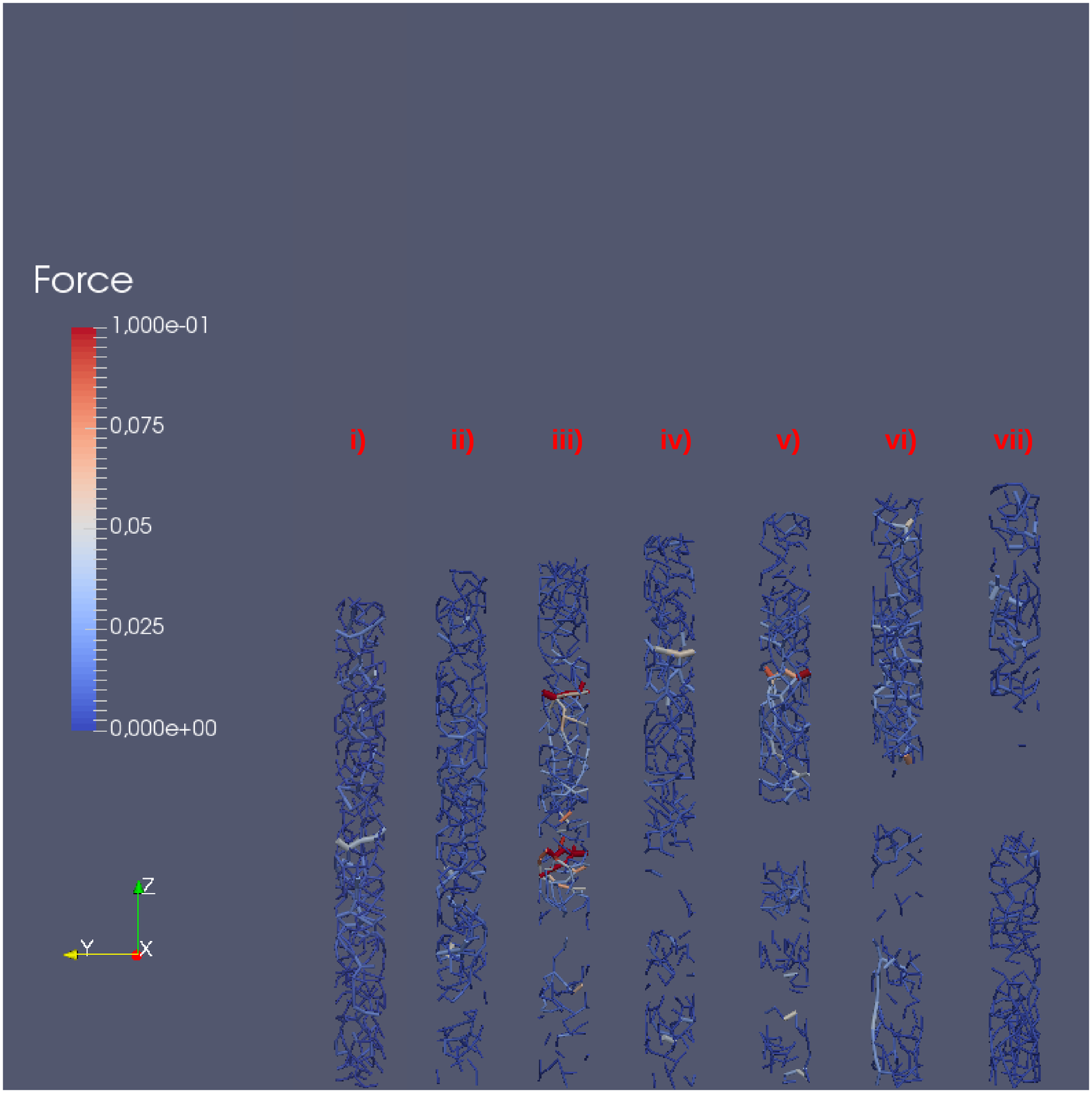}\\
 			(a)
 		\end{tabular}
 	\end{minipage}
 	\hfill
 	\begin{minipage}{0.5\linewidth}
 		\begin{tabular}{c}
 			\includegraphics[width=0.90\linewidth]{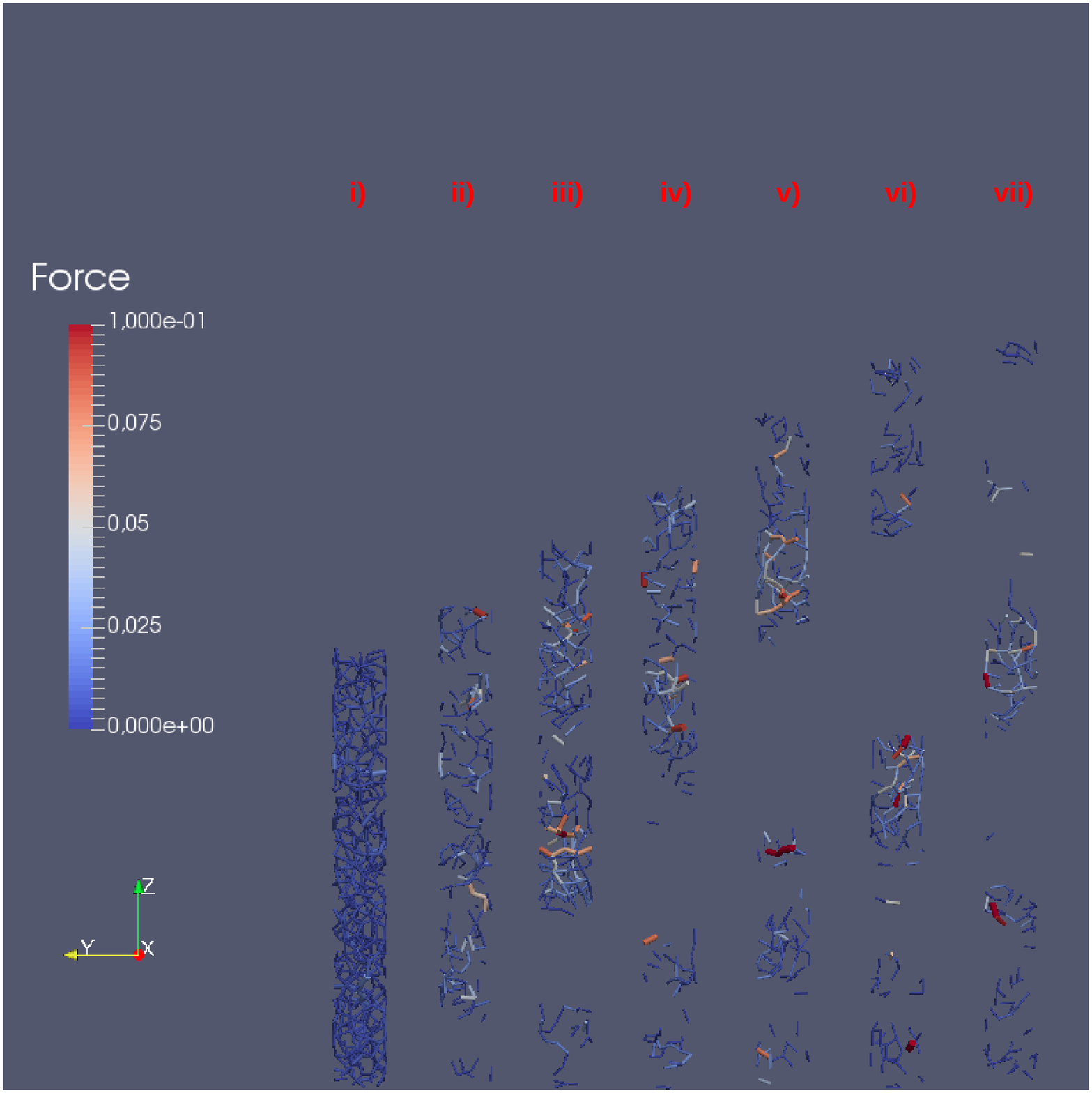}\\
 			(b)
 		\end{tabular}
 	\end{minipage}
	\hfill
	\caption{Instantaneous snapshots of the network of contact forces: (a) $N$ = 500 and $\overline{U}$ = 0.137 m/s; (b) $N$ = 500 and $\overline{U}$ = 0.164 m/s. The corresponding times are: (i) 0 s; (ii) 1 s; (iii) 1.5 s; (iv) 2 s; (v) 2.5 s; (vi) 3 s; (vii) 3.5 s. Values are in N.}
	\label{fig:num_contact_chains500}
 \end{figure}

Figs. \ref{fig:num_contact_chains250}, \ref{fig:num_contact_chains400} and \ref{fig:num_contact_chains500} present instantaneous snapshots of the network of contact forces for the 250, 400 and 500 particles beds, respectively, for both flow rates. The corresponding times, that are in the caption of figures, are the same as those in Figs. \ref{fig:num_simulations_grains250} to \ref{fig:num_simulations_grains500}. Figs. \ref{fig:num_contact_chains250} to \ref{fig:num_contact_chains500} show that the network of contact forces within plugs extend through the entire pipe cross section. This shows the importance of arches and the influence of walls for the formation of granular plugs.

In summary, in both experiments and numerical simulations granular plugs and liquid bubbles appeared. Those forms occupied the entire tube cross section and were nearly one dimensional, propagating upwards with characteristic lengths and celerities. Our numerical simulations captured well the dynamics of the liquid fluidized bed in very narrow tubes. The simulations were made with a CFD-DEM approach, where for the liquid we used the two-phase flow equations with the PIMPLE method and the big particle void fraction model, and for the solid particles we used the HSD model and considered the virtual mass and lubrication forces. The results are promising, and the methods employed in this work can be applied to more complex scenarios found in industry.

\section{Conclusions}

This paper investigated experimentally and numerically the dynamics of granular plugs in water fluidized beds in a very narrow tube. The confinement created by the narrow tube leads to the formation of alternating high- and low-compactness regions, known as plugs and bubbles, which have characteristic lengths and celerities.

In the present study, fluidized beds were formed in a 25.4 mm-ID tube and consisted of alumina beads with 6 mm diameter and specific density of 3.69 fluidized by water flows. The ratio between the tube and grain diameters was 4.23, which is considered a very narrow case. In our experiments, the fluidized bed was filmed with a high-speed camera, and the plugs were automatically identified and tracked along images by using numerical scripts. For the numerical part, we performed three-dimensional simulations using a coupled CFD-DEM code, together with numerical scripts to identify and track the granular plugs. Our numerical results using CFD-DEM were able to capture the formation of plugs under water, with the same dynamics observed in experiments.

We obtained the length scales and celerities of the granular plugs in very narrow tubes from both experiments and numerical simulations. Under the experimental conditions, granular plugs and void regions occupying the entire tube cross section were observed in the fluidized bed. Despite the small variation in water superficial velocities when compared to the variation in the number of particles, the lengths of plugs strongly depended on the water flow and were independent of the initial height of the bed. Bed celerities showed variations with both the initial bed height and water flow. The same behavior was observed in the numerical simulations.

We found from our experiments that the lengths of plugs were around 12$d$ for $\overline{U}$ = 0.137 m/s and 7$d$ for $\overline{U}$ = 0.164 m/s, while our numerical simulations showed that they were around 16$d$ for $\overline{U}$ = 0.137 m/s and 9$d$ for $\overline{U}$ = 0.164 m/s. The bed celerities showed roughly the same values for both experiments and numerical simulations: celerities in the upward direction varied between 0.012 and 0.069 m/s and in the downward direction between -0.019 and -0.115 m/s.

The results of simulations and experiments presented differences of the order of 10 \% between them with the exception of plug lengths, which presented differences of the order of 20 to 30 \%. These differences are the result of the discrete nature of plugs, that consisted of large particles (the particle diameter was 10 \% of the plug length), together with the different nature of data obtained from experiments and simulations, which were processed in different manners.

From the numerical results we computed the network of contact forces for all the tested conditions. The evolution of the contact network showed the presence of chains extending through the entire tube cross section within the granular plugs, proving the importance of arches and the influence of walls for the formation of granular plugs.

Taking into account the very discrete nature of plugs and the differences in data processing, the agreement between experiments and simulations is good. These results are encouraging: (i) we found, for the first time, the scales of the plug pattern; (ii) the numerical methodology used can be applied to more complex industrial problems.


%

\section*{Acknowledgments}
 
Fernando David C\'u\~nez Benalc\'azar is grateful to FAPESP (grant no. 2016/18189-0), and Erick de Moraes Franklin would like to express his gratitude to FAPESP (grant no. 2016/13474-9) and CNPq (grant no. 400284/2016-2) for the financial support they provided.






\bibliographystyle{elsarticle-num}
\bibliography{references1}







\end{document}